\documentclass[aip,rsi,preprint,amsmath,amssymb,]{revtex4-1}

\usepackage[latin1]{inputenc}
\usepackage{graphicx}% Include figure files
\usepackage{dcolumn}% Align table columns on decimal point
\usepackage{bm}% bold math

\begin{document}

%\preprint{}

\title{Readout for intersatellite laser interferometry: Measuring low frequency phase fluctuations of HF signals with microradian precision} %Title of paper



\author{Oliver Gerberding}
\email[]{contact@olivergerberding.com}
%\thanks{}
\altaffiliation{Present address: National Institute of Standards and Technology, Gaithersburg, Maryland 20899, USA \& Joint Quantum Institute, University of Maryland, College Park, Maryland 20741, USA}
\author{Christian Diekmann}
\altaffiliation{Present address: SpaceTech GmbH, Seelbachstraße, 88090 Immenstaad am Bodensee, Germany }
\author{Joachim Kullmann}
\author{Michael Tröbs}
\author{Ioury Bykov}
\author{Simon Barke}
\author{Nils Christopher Brause}
\author{Juan Jos\'{e} Esteban Delgado}
%\altaffiliation{Present address: Coherent Inc. , Seelbachstraße, 88090 Immenstaad am Bodensee, Germany }
\author{Thomas S. Schwarze}
\author{Jens Reiche}
\author{Karsten Danzmann}
\affiliation{Max Planck Institute for Gravitational Physics, and Institute for Gravitational Physics of the Leibniz Universität Hannover, Callinstrasse 38, 30167 Hannover, Germany}
\author{Torben Rasmussen}
\author{Torben Vendt Hansen}
\author{Anders Engaard}
\affiliation{Axcon ApS, Diplomvej 381, DK-2800 Kgs. Lyngby, Denmark}
\author{Søren Møller Pedersen}
\affiliation{DTU Space, National Space Institute, The Technical University of Denmark, Elektrovej 327, DK-2800 Kgs. Lyngby, Denmark}
\author{Oliver Jennrich}
\author{Martin Suess}
\author{Zoran Sodnik}	
\affiliation{European Space Research and Technology Centre, European Space Agency, Keplerlaan 1, 2200 AG Noordwijk, The Netherlands}
\author{Gerhard Heinzel}
\affiliation{Max Planck Institute for Gravitational Physics, and Institute for Gravitational Physics of the Leibniz Universität Hannover, Callinstrasse 38, 30167 Hannover, Germany}

\date{\today}

\begin{abstract}
%Intersatellite laser interferometers are the prime technology for space-based gravitational wave detection \cite{Danzmann2013} and for future satellite geodesy missions like GRACE Follow-On \cite{Sheard2012}.  
Precision phase readout of optical beat note signals is one of the core techniques required for intersatellite laser interferometry. Future space based gravitational wave detectors like eLISA require such a readout over a wide range of MHz frequencies, due to orbit induced Doppler shifts, with a precision in the order of $\mu \textrm{rad}/\sqrt{\textrm{Hz}}$ at frequencies between $0.1\,\textrm{mHz}$ and $1\,\textrm{Hz}$. In this paper, we present phase readout systems, so-called phasemeters, that are able to achieve such precisions and we discuss various means that have been employed to reduce noise in the analogue circuit domain and during digitisation. We also discuss the influence of some non-linear noise sources in the analogue domain of such phasemeters. And finally, we present the performance that was achieved during testing of the elegant breadboard model of the LISA phasemeter, that was developed in the scope of an ESA technology development activity. 
\end{abstract}

%\pacs{1234;5678;}% insert suggested PACS numbers in braces on next line
\keywords{phase measurement, laser interferometry, instrumentation \\
Submitted to Review of Scientific Instruments}
\maketitle %\maketitle must follow title, authors, abstract and \pacs

\section{Introduction}

Inter satellite laser interferometers are studied for space based gravitational wave detectors, like LISA, eLISA \cite{Danzmann2013}, and for satellite based geodesy missions, like GRACE Follow-On \cite{Sheard2012}. The orbit configurations of these missions lead to relative velocities of the spacecrafts, in the order of 10\,m/s, that cause Doppler shifts of the laser beams that are transmitted between any two satellites. To accommodate these, a heterodyne detection scheme using readout systems with sufficiently large bandwidth, in the order of 25 \,MHz (at a laser wavelength of $\lambda_0=1064$\,nm) is the current baseline technique. Digital phase readout systems using all digital phase-locked loops (ADPLL), similar to GPS receivers, have been developed and utilised in various laboratories \cite{Shaddock2006, Gerberding2013, Francis2014, Yu2014, Liu2014,Liang2015}. 

The most critical aspect of these readout systems, also called phasemeters, is their phase noise performance over long time scales (1000\,s), i.e. mHz Fourier frequencies. Especially the requirements for a future LISA like mission are very stringent. A noise floor below $2\pi\,\mu$rad$/\sqrt{\textrm{Hz}}$ has to be achieved at frequencies between 5\,mHz and 1\,Hz, with a relaxed performance requirement below 5\,mHz (described by a noise shape function NSF$(f)$ ). This corresponds roughly to a precision of 1\,ppm of the laser wavelength and translates to 1\,pm$/\sqrt{\textrm{Hz}}$ of effective optical path length change noise floor. % ($\lambda=1064$\,nm).  

\begin{figure}
	\includegraphics[width=\columnwidth]{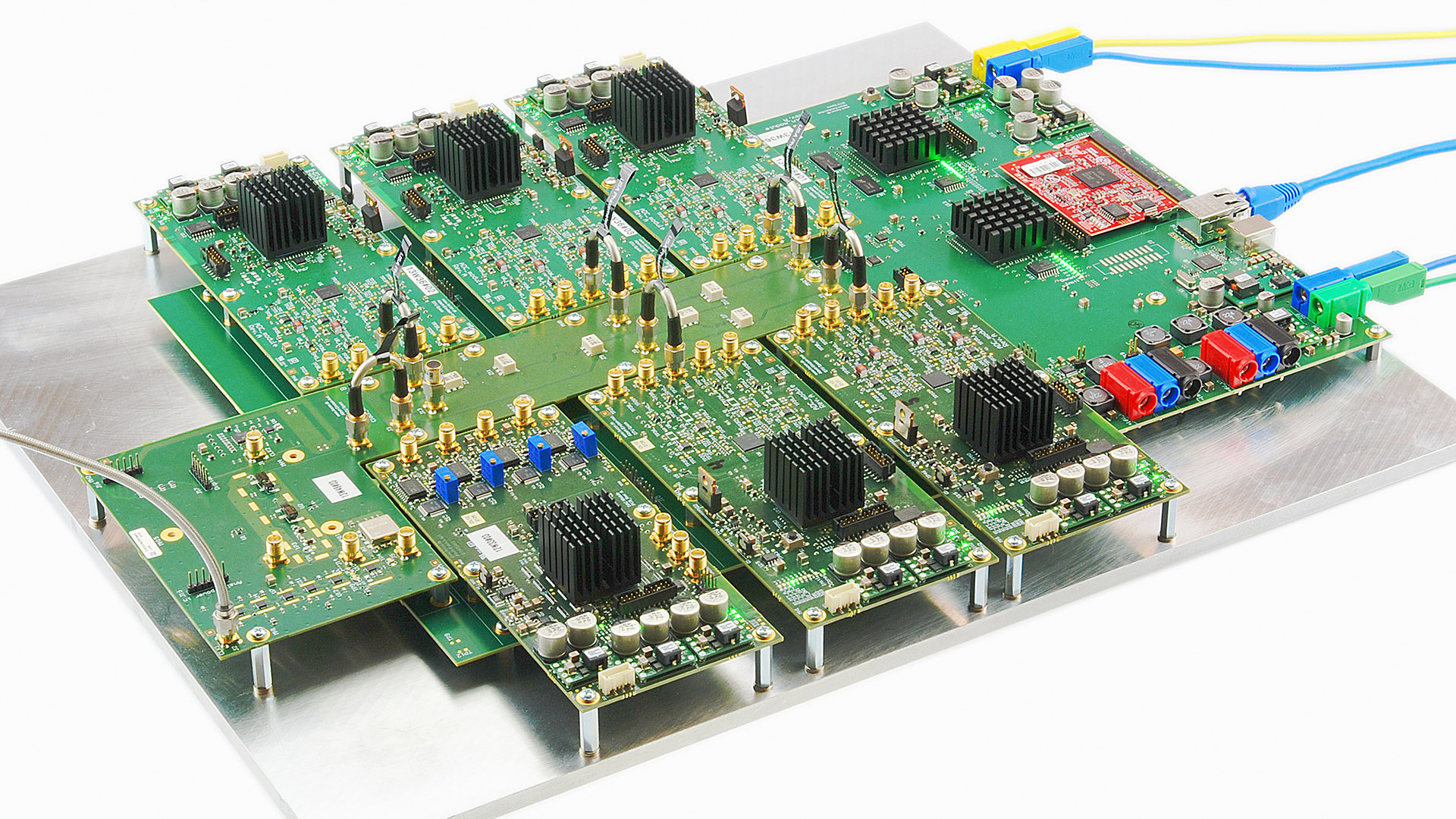}
	\caption{\label{phasemeter-full-on-small} Photograph of the assembled elegant breadboard model of the LISA phasemeter. The system consists of one main board, one clock module, one DAC module and up to five ADC modules \cite{Gerberding2012}. A total of 20 readout channels, 4 analogue control outputs, eight FPGAs and an ultra-stable clock and pilot tone distribution are included.}%
\end{figure}

We have constructed a number of phasemeter prototypes that achieve such noise levels. This in particular includes a full breadboard model of the LISA phasemeter, developed in the scope of an ESA technology development activity \cite{Gerberding2012,Barke2014}, shown in Figure \ref{phasemeter-full-on-small}. Each of these prototypes utilises a distinct implementation of the analogue part of the phase readout chain, which is one of the most critical and challenging elements of the metrology. Due to our investigations we have gained detailed insight into the relevant effects and in this article we report our current best understanding of them, together with experimental results that demonstrate the desired performance levels.

\subsection{Phase readout chain overview}

\begin{figure}
	\includegraphics[width=\columnwidth]{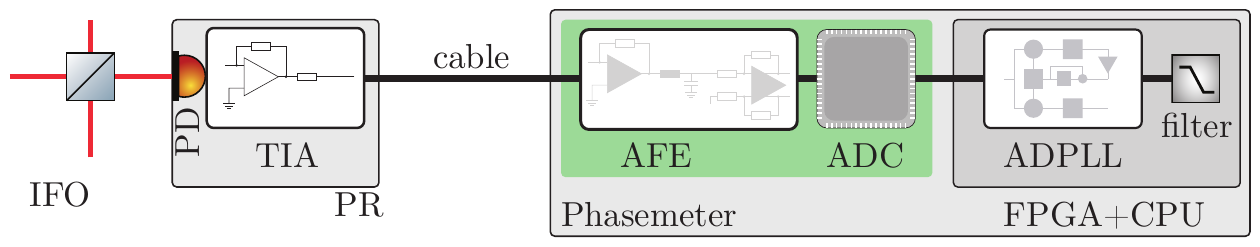}
	\caption{\label{fig:phase_meas_chain} Schematic overview of the phase readout chain. This article focuses on the elements marked green. The interferometer (IFO) signal is detected by a photo receiver (PR), which consists of a photo diode (PD) and a trans-impedance amplifier (TIA). The phasemeter receives and conditions the incoming signals in the analogue-front end (AFE) and then digitises it in the analogue-to-digital converter (ADC). A combination of Field programmable gate arrays (FPGA)s and a central processing unit (CPU) extracts the signal phase via an all-digital phase-locked loop (ADPLL) and post processes the data for storage or further use.}%
\end{figure}

To set the stage for our investigation we describe here shortly the core metrology chain for the phase readout in intersatellite interferometry, sketched in Figure \ref{fig:phase_meas_chain}. For readability we omit an introduction of the optical set-ups, which can be found in various related publications \cite{Sheard2012,Trobs2012,Spero2011}.

The starting point of our description is an optical beat note with a frequency between about $1\,$MHz and $25\,$MHz (the exact numbers depend on the individual mission and its design) that is sent to a photo diode (PD). The frequency and phase of this beat note contain the desired information about length and attitude fluctuations (by differential wavefront sensing, DWS)\cite{Schutze2014} sensed by the interferometer. The PD is usually integrated with a trans-impedance amplifier (TIA), that converts the photo current into a voltage, and, together, they form an active photo receiver (PR) \cite{Cervantes2011}. These devices are placed either directly in the interferometer or close by. Their outputs are usually single-ended analogue signals that are fed through coaxial cables to the phasemeter. %50\,$\Omega$

The phasemeter can, conceptually, be split into two parts: First, the signal conditioning in the analogue front-end (AFE) and the signal digitisation in the analogue-to-digital converter (ADC). Second, the deterministic digital signal processing (DSP) performed in FPGAs and CPUs that consists mainly of ADPLLs, appropriate filtering, decimation and signal combination.
 
The DSP units inside the phasemeter have already been described and investigated in detail \cite{Shaddock2006,Gerberding2013} and will not be discussed further in this article. Based on this earlier work, we assume that the 
noise in the measurements shown below is 
%measurement effects presented in this article are
not caused by artefacts in the DSP. However, for some of the later described, not-yet understood effects, an influence of the DSP can not be fully excluded at the current time.

The classification of the AFE as part of the phasemeter is arbitrary. All of the components can, in principle, also be part of the photo receiver or any other separate element in between the actual photo diode and the analogue-to-digital converter. The described analogue circuit effects are also present in the PR, however, they are often investigated separately, because PR are required in higher numbers, they have heritage from commercially available components and they operate in a different environment, close or on the optical benches, with, for example, better thermal stability.
The concept of building independent PRs, that send single-ended signals to the phasemeter through an impedance matched cable has been widely used and is the basis for the discussions in this article. Alternatives that use differential signal distribution, digitisation directly on the photo receiver, and stand-alone analogue-front-ends that are separated physically from the digital components of the phasemeter are possible but were discarded so far, due to concerns regarding thermal design, power consumption and overall complexity. The strong influence of temperature fluctuations, described later in this article, might lead to a revision of these design principles for future studies and implementations.
%At this point we want to remind the reader that the concept of building independent PR, that sends a single-ended signal to the phasemeter through an impedance matched cable, is, even though it is widely used, only one of many possible implementations. 
%
\subsection{Article structure}
%
%Conceptual
%	ADC
%	AFE
%
%Protoype experiments
%
%Conceptual non-linear noise sources
%	crosstalk
%	reflections
%
%Results from commercial phasemeter
We begin our review with a conceptual discussion of the signal digitisation in the ADC (Section \ref{dig_effects}). This is necessary to understand the requirements that drive the design of the AFE. The derived structure of the AFE is described next, together with a set of known noise sources (Section \ref{analog_eff}).  We then discuss non-linear effects in the AFE that can also limit the performance (Section \ref{sec:non_lin}).

The standard test procedure and a set of experiments using a combination of commercial hardware, and in-house produced components is presented to demonstrate successful implementations that were, however, limited in input frequency range. (Section \ref{sec:perf_proto}).

Next, we present the results of phase noise investigations of a breadboard model of the LISA phasemeter in Section \ref{sec:lisapms}, where realistic, LISA-like analogue input signals were used. This is accompanied with an overview of an active temperature stabilisation that was implemented to achieve, for the first time, full performance for signal frequencies of up to 25\,MHz.

%Finally, we summarize our findings and discuss shortly future experiments for fur
%Which is followed by a conclusion in which we summarize the current understanding 

%In the following we will focus almost purely on the  

%F  or the purpose of this article we divide the phase readout chain into three separate entities, the photodiode, the analogue front-end (AFE) and the digital phasemeter core and backend. This article will almost solely focus on the effects relevant in the AFE and  	
%In this article we divide the phase readout chain into three 

%Optical signal is converted from current to voltage, then digitised and then fed into digital phase-locked loop, a principle also utilised for the phase readout in GPS receivers.
%\label{}

% %\cite{Liu2014}

%\cite{Francis2014}

%\cite{Schutze2014}

%\cite{Barke2014}

%\cite{Trobs2012}

\section{Analogue-to-digital conversion}
\label{dig_effects}

The interferometer signals need to be digitised with sufficient precision and rate. The two most critical sampling parameters are the sampling frequency $f_s$ and the ADC bit depth N. 

\subsection{Truncation noise}
Quantisation noise induced by digitisation is often modelled as white additive noise, an assumption that, given the complex, noisy input signal characteristics expected in LISA like missions, is assumed to be appropriate \cite{Wannamaker2003}. The effective phase noise level due the quantisation noise depends not only on the sampling parameters \cite{Shaddock2006,Liu2014}, but also strongly on the ratio between the signal peak amplitude $v_{\textrm{s}}$ and the maximum ADC input voltage range $v_{\textrm{ADC}}$. The effective phase noise $\widetilde{\varphi}_{\textrm{trunc}}$, due to truncation, can be written as \cite{Gerberding2013}
\begin{equation}
	\widetilde{\varphi}_{\textrm{trunc}} =  \frac{v_{\textrm{ADC}}}{v_{\textrm{s}}} \frac{2^{-N}}{\sqrt{3 f_s}}.
\end{equation}
Here, the largest possible signal before the occurrence of clipping and lowest phase noise corresponds to $ v_{\textrm{s}} = v_{\textrm{ADC}}/2$.
Truncation, therefore, induces the need for proper signal gain conditioning in the AFE. The effective levels during digitisation should be as high as possible to minimise truncation noise influence, but they should also be small enough to avoid clipping. One should note, that the corresponding maximum signal calculation  has to take additive input noise and additional tones into account.
\subsection{Aliasing}
During the sampling process any signals above the Nyquist frequency ($f_s/2$) are folded into the baseband. This does not only apply to coherent signals, but also to noise. This is a critical factor, since it means that additive noise sources, like, for example, photon shot noise, that have a bandwidth larger than $f_s/2$, will accumulate after digitisation in the base band. However, these influences can be calculated and used to derive required suppressions of an anti-aliasing filter (AAF) in the AFE. A higher sampling rate reduces not only digitisation noise, but it also increases the band of frequencies that is not aliased into the signal bandwidth, simplifying the filter construction. Since faster sampling is usually accompanied by higher power consumption and other system design demands a trade-off has to be made. The response of the PR has to be taken into account as well, which will often already provide some level of suppression at higher frequencies. 

\subsection{ADC sampling jitter and pilot tone}

During the digitisation the signal is sampled using the provided ADC clock. Any phase change in the distribution or application of this clock to the sampling process causes an effective timing jitter and a corresponding phase change in the measured signals. The phase error amplitude spectral density $\widetilde{\varphi}_{\widetilde{\tau}}$ ($[\widetilde{\varphi}_{\widetilde{\tau}}] = \textrm{rad}/\sqrt{\textrm{Hz}}$) scales with the input signal frequency and the timing jitter spectral density $\widetilde{\tau_{\textrm{}}}$ ($[\widetilde{\tau_{\textrm{}}}] = \textrm{s}/\sqrt{\textrm{Hz}}$):
 \begin{equation}
\widetilde{\varphi}_{\widetilde{\tau}} = 2 \pi f_s \cdot \widetilde{\tau_{\textrm{}}}.
 \end{equation}
Experiments in the last decade have revealed that all suitable ADCs have an inherent $1/f$ timing jitter noise that prevents phasemeters from directly achieving the desired performance levels. This noise is inherent to the ADCs themselves, and is present even with a perfect clock distribution. % it can not be suppressed sufficiently.

However, the timing jitter can be accounted for by utilising a reference tone \cite{Shaddock2006}, also called pilot tone, that is fed into all ADC channels. The phase deviations of this tone, whose frequency $f_{\textrm{p}}$ is well known, can be converted into an effective measurement of the timing jitter. The corrected phase for each input channel $\varphi_{i,c}$ is given by the combination of the signal phase measurement $\varphi_{i,s} $ and the individual pilot tone measurement $\varphi_{i,p} $:
\begin{equation}
\varphi_{i,c} = \varphi_{i,s}- \varphi_{i,p} \frac{f_{i,s}}{f_p}.
\end{equation}
To implement such a correction scheme the AFE needs to be able to distribute the pilot tone, to add it to each individual signal and to deliver both signals after appropriate signal conditioning to the ADC, without spoiling the phase noise performance of either tone. The requirement for the pilot depends on the frequency ratio.

In a timing picture the pilot tone is the effective, jitter free (reduced) reference clock. 

\subsection{ADC analogue parameters}
So far we have focused on the sampling parameters of ADCs. However, the analogue effects described in the following sections \ref{analog_eff} and \ref{sec:non_lin} are also relevant for the ADCs. We have empirically tested candidate ADCs for compliance by directly performing pilot tone corrected phase noise measurements in relevant thermal environments, as described in Section \ref{sec:perf_proto}. 
%with  for their In general, we ensure that the ADCs are compliant by directly measuring their phase noise performance. 

\section{Analogue front-end}
\label{analog_eff}

The signal conditioning for digitisation is done in the AFE. In addition to the above described functions the AFE also converts the single ended input signal into a differential signal with a desired gain and impedance to match the following ADC.
%which is required for all of the considered and utilised ADCs. 
The separate functions required of the AFE are listed in the following:
%The now defined functions of the AFE are listed in the following:
\begin{itemize}
	\item{Buffer: Receive PR signal from coaxial cable.}
	\item{Amplifier: Amplify signal with desired gain.}
	\item{Anti-aliasing filter: Reduce unwanted signal components above the Nyquist frequency. }
	\item{Pilot tone distribution: Deliver the pilot tone to individual AFEs without introducing excess phase noise.}
	\item{Adder: Add the pilot tone to the PR signal before digitisation.}
	\item{Single-ended to differential converter: Prepare the signal for the ADC.}
\end{itemize}

The critical effects in the AFE that contribute to phase noise are summarised below.

\subsection{Additive noise sources}
Any component in the AFE can contribute additive electronic noise. The induced corresponding phase noise is simply given by the ratio of the noise and the root-mean-square (RMS) signal amplitude at the given point in the measurement chain \cite{Gerberding2013}. Hence, the most critical point for introducing noise is the one with the smallest signal. For LISA this is generally the first amplification in the trans-impedance amplifier of the photo receiver \cite{Cervantes2011}. Signals sent to the phasemeter are assumed to be amplified to sufficiently high levels, such that fundamental noise from passive components, like Johnson noise from resistors, is sufficiently small.

Operational amplifiers and other active components used in the AFE can also introduce voltage noise in the MHz signal band. To achieve effective phase noise levels below $6\,\mu$rad/$\sqrt{\textrm{Hz}}$ these contributions should be below 600\,nV/$\sqrt{\textrm{Hz}}$ for typical minimal signal RMS amplitudes of 100\,mV. Commercially available low-noise operational amplifiers achieve noise levels on the order of a few nV/$\sqrt{\textrm{Hz}}$, which makes this contribution non-critical for the circuits used in the AFE, like inverting-amplifiers with moderate levels of gain (the noise coupling of operational amplifiers is gain dependent \cite{Carter2003}). % The coupling of the operational amplifier noise   %choice of components  which enables us to take this contributions into account by simply choosing compliant components.  %  for adding noise in the detection scheme is In general, however, the LISA metrology uses a significantly large signal amplification in the PR to ensure that noise added in the AFE is sufficiently small in comparison to the signal amplitude. While noise introduced by passive components (like Johnson noise in resistors), is fully negligible, the noise of active components, like operation amplifiers, can be taken into account based on available specifications. While care has to be taken, this influence is not considered critical for the AFE.

\subsection{Phase noise}
Changes of the signal phase in the AFE directly spoil the measurement performance. Each component in the AFE can influence the phase of a transmitted signal and here we shortly describe two types of phase noise coupling that are relevant and occurred during our experiments. %two relevant t and such effects, occurring within the measurement band, have to be suppressed or controlled. %Any Unwanted change in this phase influence in the measurement band can cause excess phase noise, which can poil the final measurement performance. 

\subsubsection{Temperature coupling through finite phase response}
Components in the AFE have non-flat transfer functions $T(f)$. Any frequency dependent phase response, as, for example, induced by a simple signal delay, influences the phase propagation through the measurement chain. Constant, linear frequency dependent phase response is not critical for LISA, because the signal frequency changes are happening at frequencies below the measurements band ($<1\,$mHz). If significant frequency drifts occur within the measurement band, like expected for GRACE Follow-On \cite{Sheard2012}, the overall phase response either has to be sufficiently flat, or a correction in post-processing could be applied. 

Even for a constant input frequency $f_s$ the phase response can cause significant phase noise if its value, $\arg{(T(f_s))}$, is influenced by in-band fluctuations of temperature, for example due to the change of a resistance or capacitance. Flat transfer functions are in general less susceptible to these fluctuations, as the susceptibility to thermal variations depends on the absolute slope of the phase response. Accordingly, the phase of signals at higher frequencies is more susceptible to thermal changes in bandwidth limited circuits. A reduction of temperature fluctuations is one way to mitigate the induced phase noise (as we present in Section \ref{sec:lisapms}), but limitations in size, power consumption and cost of the apparatus limit this approach. Performing detailed analysis of each components susceptibility to temperature changes is required to reduce the coupling itself. Choice of more stable components and thermal compensation are some of the methods that can be applied. A rule of thumb that we applied during our investigations is to use components with as large a bandwidth as possible. This approximates flat transfer functions responses that minimise the coupling to thermal effects. 

Due to this coupling the most critical component in the AFE is the anti-aliasing filter. It requires a large suppression above the Nyquist frequency to ensure shot-noise limited performance (see Section \ref{dig_effects}), which strongly limits its bandwidth and phase response flatness, and it has to have sufficiently low coupling of thermal noise into phase noise. 
Our investigations presented here do not include a detailed analysis of, nor a design for, the AAF. Based on our later presented results it is assumed, however, that integrating the AAF into the PRs might be beneficial, due to the better thermal stability.  

\subsubsection{Flicker phase noise}
Another source of phase noise is flicker noise, that is, for example, induced by a $1/f$ noise at DC in amplifiers that effectively modulates the carrier phase \cite{Boudot2012}. In our experiments we have avoided using components that contribute any such excess noise, by performing empirical tests. The investigation of such effects is a separate topic that goes beyond the work presented here and we found no essential components that are limited by them. Frequency mixers are a prominent example of components limited by such noise and we have observed it in earlier studies \cite{Gerberding2013}.

\subsection{Pilot tone distribution}
The use of a pilot tone requires it to be distributed to each analogue measurement channel. This involves impedance matched signal splitting, signal routing, and, as discussed later in Section \ref{sec:non_lin}, isolation between channels. The phase noise analysis done for the AFE also has to be applied to the pilot tone distribution (PTD), being, however, slightly simplified because the pilot has a single, fixed frequency. 

\section{Non-linear effects}
\label{sec:non_lin}

The phase measurement effects described above are critical for understanding and designing phasemeter systems. So far we have described the influence of processes that either cause an additive noise or a phase noise of the coherent signal. Both of these noise types cause an effective phase noise (additive noise is converted into phase noise in the mixer in the ADPLL), whose underlying processes are linear with respect to the noise amplitude within the bandwidth of the ADPLL.  
%which is linear with respect to the underlying processes. 
But, other types of interactions can also limit the achievable performance. The one we consider here is the presence of additional, parasitic tones at the signal frequency or close by. The ADPLL does not distinguish between the actual signal tone and the parasitic one and will act as if a single disturbed tone is present (parasitic tones outside the ADPLL bandwidth are rejected).

A simple way to determine the combined, disturbed tone is to describe both tones as phasors in the complex plane at the readout frequency. The two phasors, or 2-dimensional vectors, each defined by an amplitude and a phase, are added, resulting in the effective phasor, with a corresponding phase and amplitude. The difference between the effective phase and the phase of the signal tone is the phase error. 
The vector addition causes the coupling into the measurement of phase or amplitude changes of the parasitic tone relative to the signal tone to be highly non-linear. A parasitic tone that has a constant phase and amplitude ratio relative to the signal, will only result in a DC phase error, which is anyhow neglected for longitudinal displacement measurements (not for DWS though).

\begin{figure}
	\includegraphics[width=\columnwidth]{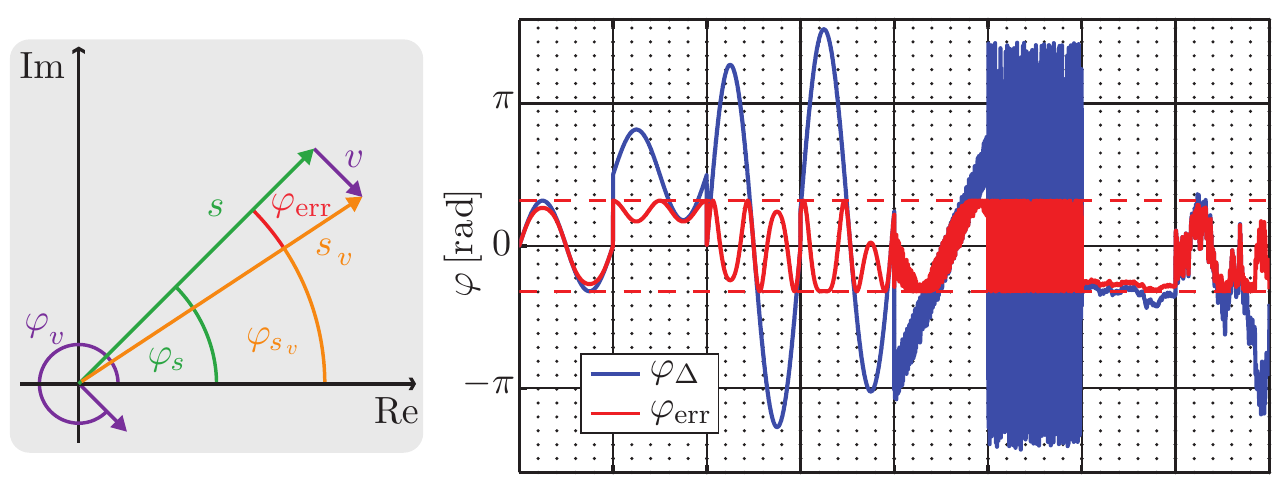}
	\caption{\label{crosstalk_sketch_ts} Illustration of small vector noise coupling. The left side shows a sketch of the vector addition responsible for the non-linear noise coupling in case of small vector noise. The right side shows a simulated time series of a small vector phase modulation for coherent modulations with different DC offsets and amplitudes and for white noise and $1/f$ noise.}%
\end{figure}
The parasitic signals become more relevant when they have non-negligible dynamics relative to the signal tone. Such unwanted, parasitic tones are made as small as possible in comparison to the signal in the system design. Their effect is illustrated in Figure \ref{crosstalk_sketch_ts}. In this case, a small phasor, or vector, $\vec{v}_{\textrm{v}}$ is added to the larger, dominating signal tone $\vec{v}_{\textrm{s}}$. The phase noise caused in such a situation is often referred to as small-vector noise. The complex coupling of relative phase fluctuations ($\varphi_{\Delta} = \varphi_{\textrm{s}} -\varphi_{\textrm{v}}$) into phase error $\varphi_{\textrm{err}}$ can be approximated for the small parasitic vector to
\begin{equation}
\varphi_{\textrm{err}} = \frac{v_{\textrm{v}}}{v_{\textrm{s}}} \sin{( \varphi_{\textrm{s}} -\varphi_{\textrm{v}} )}.
\label{eq:smallvector}
\end{equation}
%
%In this case one can approximate the earlier described coupling and due to the earlier desand their phase dependent coupling lead to them being referred to as small-vector noise. 
Such non-linear noise couplings are well known in the gravitational wave community. A small vector noise has been observed and suppressed in LISA Pathfinder \cite{Hechenblaikner2013}, and stray light and ghost beams, which are present in many laser interferometers, cause a very similar effect as well.
% Hence, it will determine a wrong phase The combined input sigal The coupling of these parasitics into the phase measurement depends on their relative phase to the signal and it is, therefore, highly non-linear \cite{Hechenblaikner2013}. 

%In the approximation of a small parasitic vector in comparison to the signal ($v_{\textrm{v}}\ll v_{\textrm{s}}$), the phase error can be written as
%%
%\begin{equation}
%\varphi_{\textrm{err}} = \frac{v_{\textrm{v}}}{v_{\textrm{s}}} \sin{( \varphi_{\textrm{s}} -\varphi_{\textrm{v}} )}.
%\label{eq:smallvector}
%\end{equation}
%%
%The resulting effects depend strongly on the phase difference ($\varphi_{\Delta} = \varphi_{\textrm{s}} -\varphi_{\textrm{v}}$) variations, as illustrated in Figure~\ref{crosstalk_sketch_ts}.
In case of small relative phase dynamics $\varphi_{\Delta}$, and DC phases close to the linear regimes of Equation \ref{eq:smallvector}, the induced phase noise is linear to the dynamics, and its amplitude is attenuated by the ratio $\frac{v_{\textrm{v}}}{v_{\textrm{s}}}$. In the more general case of large dynamics and arbitrary phase relations, no simple method exists to predict the phase error from the relative dynamics. Some examples for this coupling are shown in simulated time series on the right of Figure \ref{crosstalk_sketch_ts}. Large low frequency changes can be unconverted to higher frequency errors, due to the non-linear coupling. An important feature of this noise coupling is, that the induced phase error has a maximum amplitude, which is also defined by the ratio $\frac{v_{\textrm{v}}}{v_{\textrm{s}}}$. 
This leads to a characteristic and often diagnostically useful time-domain signature of a constant width envelope.
%Even for small vector noise, no simple method exists to determine the phase error behaviour based on the phase fluctuations, because it not only depends on the phase spectra of the small vector, but also on its DC phase. However, small vector noises have the important feature that they have a maximum amplitude, given by the ratio $\frac{v_{\textrm{v}}}{v_{\textrm{s}}} $.
Small vector noise driven by low-frequency random processes with a $1/f$ spectrum, like, for example, temperature fluctuations, often show a characteristic phase noise spectrum. Such a spectrum flattens out to lower frequencies (due to the maximum error amplitude) and a characteristic noise shoulder is observed at the high frequency end of the generated noise bandwidth. If the readout bandwidth of the phasemeter is low in comparison to the generated noise, the drop-off to higher frequencies might not be observed. In this case a small vector noise can be misinterpreted as white readout noise and the underlying non-linear coupling makes eventual noise hunting more complex. 

\subsection{Cross talk}

The most relevant parasitic tones in a multi channel phasemeter are cross talk components from other channels operating at the same, or nearby frequencies. The phase of these tones can be totally unrelated, potentially causing a very dynamic behaviour of $\varphi_{\Delta}$. This is, for example, the case for the two test mass and reference interferometers on the two optical benches in a LISA satellite \cite{Trobs2013}.
The parasitic voltage in the case of cross talk can be modelled by the product of the amplitude of the unrelated signal $v_{\textrm{u}}$ with a cross talk coupling factor $C$.
Assuming a coupling as described in Equation \ref{eq:smallvector}, equal amplitudes for all input signals ($v_{\textrm{u}} = v_{\textrm{s}}$), and a performance requirement of better than $2\pi\,\mu$rad, a crosstalk suppression in the order of $C = -105$\,dB is necessary. This level is determined from a simple maximum error calculation. Better estimates can be made by applying full models of the expected phase dynamics and using them to make a prediction of the behaviour of $\varphi_{\Delta}$.

Cross talk can occur at every stage of the analogue signal processing, including, prominently, the PR \cite{Joshi2012}. Effects like electro-magnetic interference, power supply or ground stability can limit the achievable cross talk suppression between individual channels. Mitigating these effects requires detailed analysis and design of circuits, layouts, PCB stacks, component placements, electro-magnetic shielding and more. These investigations exceed the scope of this article. However, we will briefly discuss the two most critical points in the AFE of our phasemeter systems where the achievable limit of cross talk suppression is determined by component choice and simple, schematic level circuit design.
%Mitigating this effect requires a detailed simulation analysis of the AFE, including its layout and boundary conditions. But here we won't discuss it in detail, because it is an elaborate problem that far exceeded our current analysis. The AFE design can, however, rather easily take into account two of the most critical places where cross talk is introduced: 

Multi channel ADCs have inherent cross talk, often with a coupling of about -90\,dB, which is documented in their data sheets. Hence, using multi-channel ADCs for the readout of independent IFOs should be avoided to mitigate this problem. On the other hand, using them to read out the four signals of a quadrant photo diode is suitable, because these signals are already mixed due to the cross talk in the PR, and they are very similar. They are dominated by the DWS signal, keeping the value of $\varphi_{\Delta}$ small and even well measured.

The pilot tone distribution is a direct link that connects all input channels. Sufficient care has to be applied to ensure that no significant portion of the signals is transmitted back through the PTD to other signals. The adder components in the AFE are critical, they should provide sufficient initial suppression of backwards reflected signals. Additional isolating components (buffers, filters, isolators) can also be introduced into the PTD, as long as they don't contribute excess phase noise.

\subsection{Analogue reflections}

\begin{figure}
	\includegraphics[width=\columnwidth]{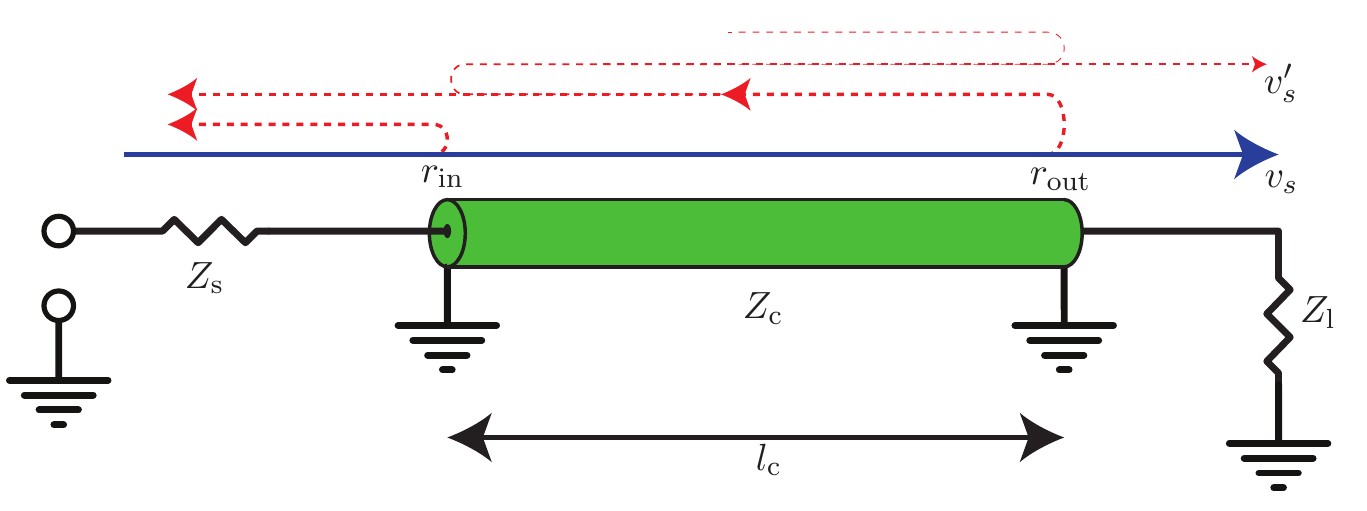}
	\caption{\label{fig:refl_analogue-up} Sketch for signal reflections occurring in the signal distribution from, for example, the photo diode to the analogue-front end. For small reflection coefficients $r_x$ a dominating parasitic signal $v^\prime_s$ is delayed by twice the cable length $l_c$ and contaminates the actual signal $v_s$.}%
\end{figure}

Small vector noise can also be caused by a tone that is contaminated with a small delayed version of itself. This can occur if signal reflections are present, for example due to impedance mismatches. Figure \ref{fig:refl_analogue-up} shows a simple model of such an effect. A signal is fed from the photo diode (the source with impedance $Z_{\textrm{s}}$) through a cable (impedance $Z_{\textrm{c}}$ and length $l_{\textrm{c}}$) to the phasemeter (the load with impedance $Z_{\textrm{l}}$). The impedance mismatches give rise to two reflection coefficients, here denoted as $r_{\textrm{in}}$ and $r_{\textrm{out}}$. For small reflectivities the amplitude of the first delayed signal $v_s^\prime$ reaching the load (we discard multiple reflections) is attenuated by $r_{\textrm{in}} \cdot r_{\textrm{out}}$. The amplitude of the actual signal $v_s$ is, however, almost undisturbed. 
The phase of $v_s^\prime$ relative to $v_s$ is determined by its dynamics and by twice the delay $\tau_{\textrm{c}}$, which is given by the cable length and the effective signal speed, in typical coaxial cables $\tau_{\textrm{c}}\approx l_{\textrm{c}}/ (2/3c)$. Using these relations and Equation \ref{eq:smallvector} we can estimate the induced phase noise by such an analogue reflection:
\begin{equation}
\varphi_{\textrm{err}}  \approx r_{\textrm{in}}  r_{\textrm{out}} \sin{( 2 \frac{d \varphi_{\textrm{s}}}{d t} \tau )} \approx r_{\textrm{in}}  r_{\textrm{out}} \sin{( 3 \frac{d \varphi_{\textrm{s}}}{d t} \frac{l_{\textrm{c}}}{c} )}.
\label{eq:smallvectoranal}
\end{equation}
%
%The coupling of such signals depends on its effective delay, its amplitude, and on the signal dynamics. 
The expected signal dynamics for inter-satellite laser interferometry are quite large (they are dominated by laser frequency noise the level of which depends on the applied stabilisation scheme). Assuming the laser frequency noise $\widetilde{f}$ is still small in comparison to the delay we can simplify the coupling,
\begin{equation}
\widetilde{\varphi}_{\textrm{err}} (f)  \approx  \frac{ 3 r_{\textrm{in}}  r_{\textrm{out}}   \widetilde{f}(f) l_{\textrm{c}}}{c} .
\label{eq:smallvectoranal2}
\end{equation}
Even for large levels of in-band laser frequency noise (3000\,Hz/$\sqrt{\textrm{Hz}}$) and cable lengths on the order of 2\,m this coupling only becomes relevant for large reflection coefficients of more than 10\%. However, if laser frequency noise levels exceed the linear range, even at much lower frequencies, effective noise levels can be upconverted to in-band Fourier frequencies, due to the non-linear coupling. Our presented model is largely simplified and aims to give insight into a possible explanation for an unexpected noise floor that we have observed in the later described measurements, which was found to be somewhat influenced by signal dynamics, cable lengths and impedance matching. Influences of our more complex signal networks, including the pilot tone distribution, and of frequency dependent impedance values have not yet been accounted for and might need to be analysed in future studies.
%Large signal dynamics, as they are expected for inter satellite interferometry, can modulate the phase between the small tone and the actual tone in such a way, that an effective noise floor is introduced. 
%Since the coupling One has to keep in mind at this point, that this coupling is non-linear, and that strong, low-frequency phase variations can, therefore, be converted up. 
%We find some evidence during our later presented measurements for such couplings.  

\section{Performance investigations using phasemeter prototypes}
\label{sec:perf_proto}
\begin{figure}
	\includegraphics[width=\columnwidth]{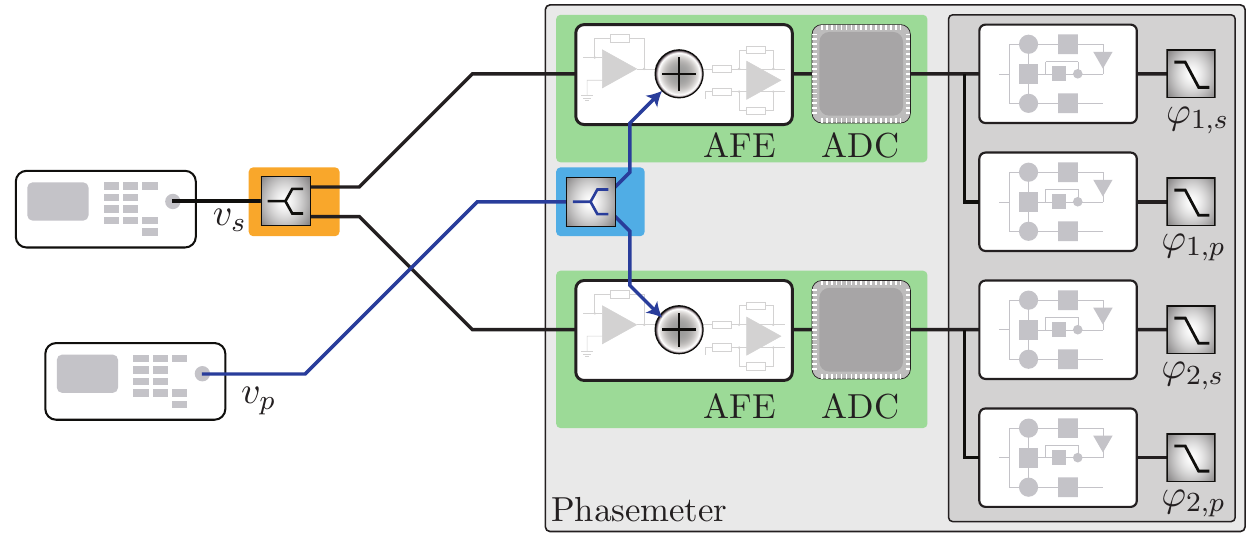}
	\caption{\label{fig:meas_set-up} Sketch of a null or zero measurement with two channels. The signal is split in a signal splitter (marked in orange) and fed into two readout channels. A pilot tone is also fed to the phasemeter, split and distributed via the PTD (marked blue) and added. Two all-digital phase-locked loops are used to read out the phase of the signal and the pilot tone for each channel.}%
\end{figure}
A standard technique for investigating the noise of phase readout systems is the so-called null or zero measurement. A single signal is split into two or more readout channels and the differences in the measurement phases are used to investigate the noise floor. Figure \ref{fig:meas_set-up} shows this principle for a two channel measurement. Throughout our investigation we performed multi-channel measurements and we plot the noise for each channel $\widetilde{\varphi}_{i}$ by computing the difference between its phase and the measured phase average: 
\begin{equation}
\widetilde{\varphi}_{i} = \varphi_{i,s}- \frac{1}{N}\sum_{k=1}^{N} \varphi_{k,s}.
\label{eq:perf_calc}
\end{equation}
This type of measurement is easily implemented and well suited to hunt critical noise sources. But, the subtraction of almost equal signals makes this scheme insensitive to common non-linear effects \cite{Gerberding2013}. Cross talk influence, for example, can not be measured well, since all channels already measure the exact same signal. Noise due to thermal fluctuations is, if it is common between all channels, also not detectable by such measurements. Common additive noise can, in the linear operating range of the ADPLL, not be detected either. The noise floor due to uncorrelated additive noise depends on the number of compared channels, as evident from Equation \ref{eq:perf_calc}.

% Since our analysis method always compares multiple channels, which leads tthe noise floor due to uncorrelated additive noise is increased by $\sqrt{2}$ in comparison to the single channel performance, if all channels are limited by roughly the same levels.
%

An important first result of our investigations is that some commercially available ADC circuits, using amplifiers for ADC signal conditioning, are capable of achieving the desired phase noise performance levels even in rather uncontrolled thermal environments. These systems only require the addition of a pilot tone to the input signals to enable the necessary timing jitter correction. We used this fact and combined an FPGA with a suitable ADC card (FMC107 from 4DSP \cite{DS_FMC107}) to evaluate the feasibility and performance of different pilot tone distribution and adding schemes.
For these measurement we used a rather stable input signal, produced by a commercial signal generator, and split and distributed it via a thermally isolated, resistive power splitter (also referred to as resistive tee). Thereby, we ensured that no significant phase noise was introduced by the signal distribution. 

\subsection{Active front-end}
\begin{figure}
	\includegraphics[width=\columnwidth]{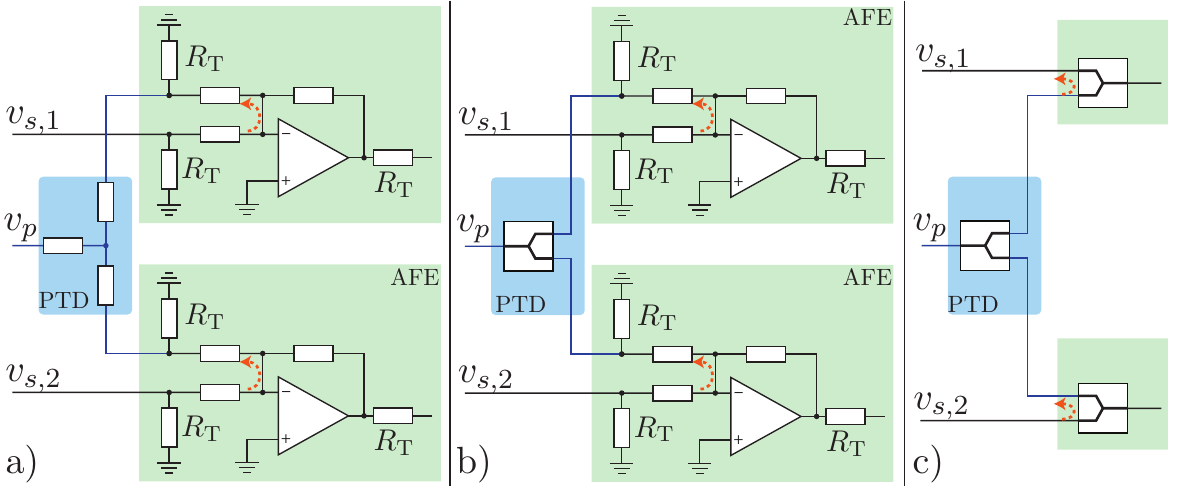}
	\caption{\label{fig:circuit_sketch} Sketch of the schematics for the tested pilot tone addition schemes, each one is shown for two channels. a)~Active pilot addition and distribution via resistive power splitter. b)~Active pilot addition and distribution via transformer based power splitter and addition via inverting adder circuits. c)~Passive pilot addition and distribution via transformer based power splitters. The orange arrows indicate signal leakage into the PTD that leads to cross talk.}%
\end{figure}

We tested two schemes using an active pilot tone adder. High-bandwidth operation amplifiers were used in simple inverting adder circuits (also referred to as summing amplifiers) to combine the signals and the pilot tone. The PTD was implemented in two ways,
once using a resistive 50\,$\Omega$ matched splitter, and once using commercial, transformer based power splitter (also referred to as hybrid coupler) with a flat amplitude response in the signal bandwidth. % Reference to schematic
Both options were integrated, together with the adders, on a separate circuit board, which was connected to the signal source and to the ADC card using coaxial cables. Sketches of the implemented schematics are shown in Figure \ref{fig:circuit_sketch}, a) and b). A passive thermal isolation of the components was added by wrapping them in bubble wrap foil.
%\begin{figure}
%	\includegraphics[width=\columnwidth]{PT_ADD_ACT_SPLIT_MC_spec.pdf}
%	\caption[Performance of active PT adder and power-splitter distribution]
%	{Performance of active pilot tone adder and power-splitter distribution. The signal frequency was at 11\,MHz, the pilot tone at 33\,MHz. Channel A was defect. The dashed lines show the performance without jitter correction.}
%	\label{fig:PT_ADD_ACT_SPLIT_MC_spec}
%\end{figure}
%

\begin{figure}
	\includegraphics[width=\columnwidth]{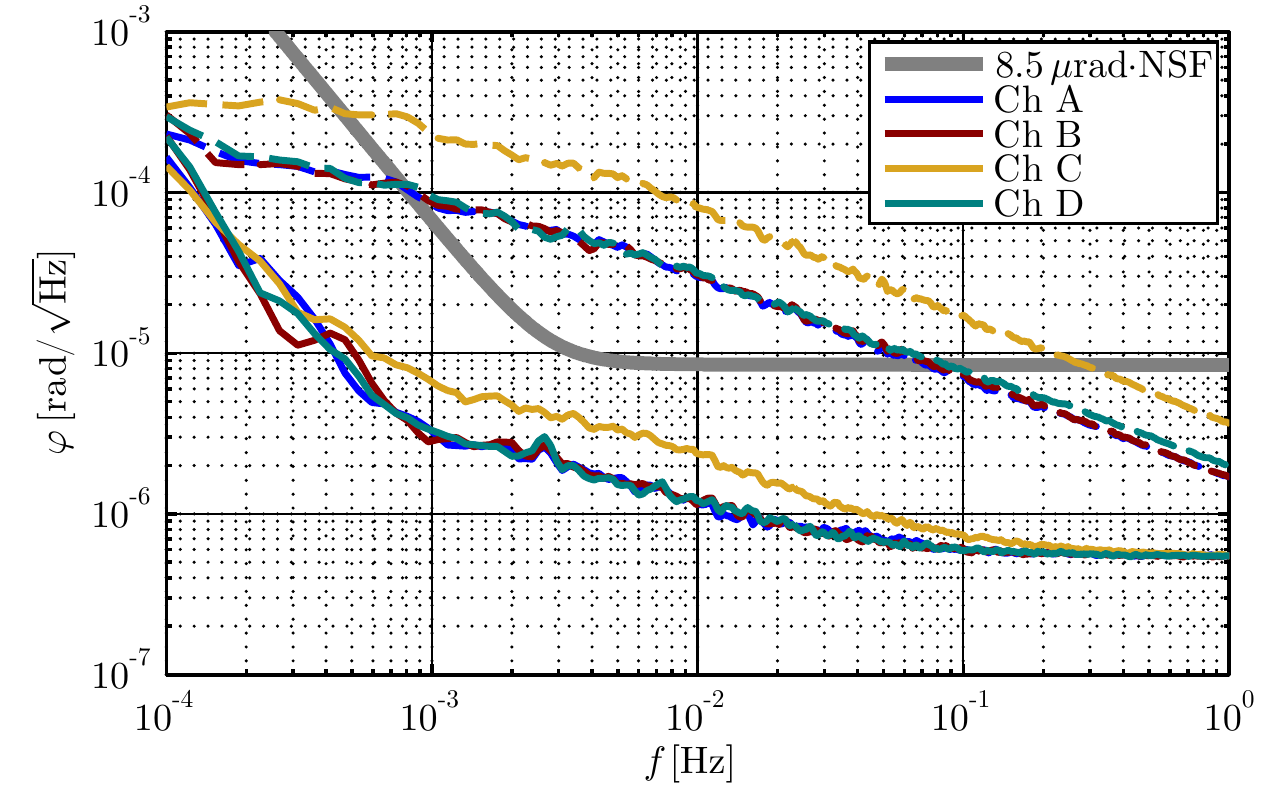}
	\caption[Performance of active PT adder and resistive distribution]
	{Performance of active pilot tone adder and resistive distribution. The signal frequency was at 11\,MHz, the pilot tone at 33\,MHz. Excess noise in C is not understood. The dashed lines show the performance without jitter correction.}
	\label{fig:PT_ADD_ACT_SPLIT_R_spec}
\end{figure}

The achieved phase noise performance levels are shown in Figure \ref{fig:PT_ADD_ACT_SPLIT_R_spec} for the resistive power splitter set-up, the levels are also representative for the once achieved with the set-up using transformer based power splitters.  The dashed lines represent the spectral density of the differences of the phase errors for the uncorrected signals. The straight lines show the same signals corrected for pilot tone jitter. Interestingly, one of the channels (C) shows an increased phase noise before correction. This is, however, sufficiently suppressed afterwards.  Both schemes were able to achieve the desired performance levels. This demonstrates that the operational amplifier based adding circuits, as well as the pilot tone distribution via resistive network or commercial power splitters are suitable to perform at the desired noise levels for the specific set of frequencies used in each measurement. 

In terms of suppression of crosstalk via the PTD the power splitter implementation is considered favourable. A cross talk signal introduced into the PT chain in one of the adders is attenuated by the isolation of the power splitters (typically 30\,dB) before it can reach another signal channel. A summing amplifier circuit provides a strong isolation between inputs, due to the virtual ground at the inverting input, which gives a strong initial cross talk suppression. However, a reduction of the nominal cross talk coupling via the PT distribution to less than 105\,dB requires additional components.

\subsection{Passive front-end}

Based on the positive results with the passive pilot tone distribution an effort was started to test a fully passive PTD and adder that uses two types of commercial, transformer based power splitters. The benefit of such an implementation is that it consumes no power and thus generates very little heat, which simplifies any thermal isolation schemes.

%\begin{figure}
%	\includegraphics[width=\columnwidth]{PT_ADD_SPLIT_MC_ADP-2-1W_spec.pdf}
%	\caption[Performance of passive PT adder and distribution with ADP-2-1W]
%	{Performance of passive pilot tone adder and distribution with ADP-2-1W. The signal frequency was at 15\,MHz, the pilot tone at 35\,MHz. The dashed lines show the performance without jitter correction.}
%	\label{fig:PT_ADD_SPLIT_MC_ADP-2-1W_spec}
%\end{figure}
\begin{figure}
	\includegraphics[width=\columnwidth]{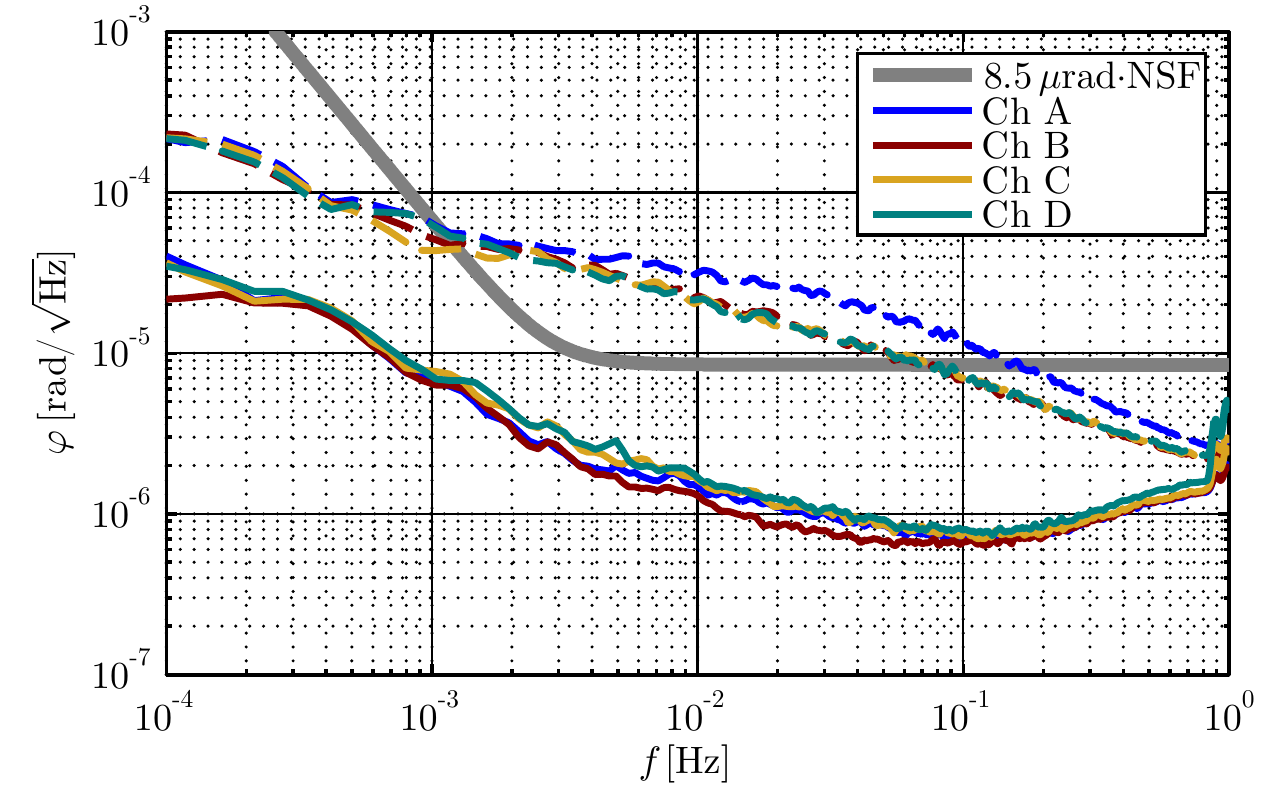}
	\caption[Performance of passive PT adder and distribution with PSC-2-1]
	{Performance of passive pilot tone adder and distribution with PSC-2-1. The signal frequency was at 15\,MHz, the pilot tone at 35\,MHz. The dashed lines show the performance without jitter correction.}
	\label{fig:PT_ADD_SPLIT_MC_PSC-2-1_spec}
\end{figure}

We have constructed AFEs with two different power splitters, one with a plastic housing (ADP-2-1, Mini-Circuits) and one in a metal housing (PSC-2-1, Mini-Circuits). Separate circuit boards containing the PT adder were, again, placed in passive thermal isolation and connected to the ADC card using impedance matched coaxial cables. A sketch of the schematics is shown in c) in Figure \ref{fig:circuit_sketch}. We investigated the phase noise performance and found that both implementations were also able to achieve the desired levels for some specific set of signal frequencies. Figure \ref{fig:PT_ADD_SPLIT_MC_PSC-2-1_spec} shows the performance for the set-up using the PSC-2-1,  which is also representative for the performance achieved with the ADP-2-1.
%(see Figure \ref{fig:PT_ADD_SPLIT_MC_ADP-2-1W_spec} and \ref{fig:PT_ADD_SPLIT_MC_PSC-2-1_spec}) 

We investigated the described schemes only with the given temperature stability and for a small set of frequencies, because the prototypes were aimed at single frequency operation. Choosing smaller (below $\approx$10\,MHz) or higher (above $\approx$20\,MHz) input frequencies  did, however, indicate the presence of excess noise. This was expected due to the bandpass characteristic of the transformer based power splitter. Their non-flat transfer function at lower and higher frequencies increases the coupling of temperature fluctuations into phase. It is currently not known, if this could have been reduced by further improved thermal stability to acceptable levels.
%The shown results could only be achieved, with the given temperature stability, for certain frequencies. The signal frequencies were chosen to be roughly in the middle of the signal splitter bandwidth, the point with the expected minimal temperature coupling. The pilot tone was at the high frequency 
%Besides the shown results we Especially higher frequencies were simply avoided during these tests, either because the prototypes were only aimed at single frequency operation, or because of time constraints.

%\begin{figure}
%	\includegraphics[width=\columnwidth]{PT_ADD_SPLIT_MC_PSC-2-1_spec.pdf}
%	\caption[Performance of passive PT adder and distribution with PSC-2-1]
%	{Performance of passive pilot tone adder and distribution with PSC-2-1. The signal frequency was at 15\,MHz, the pilot tone at 35\,MHz. The dashed lines show the performance without jitter correction.}
%	\label{fig:PT_ADD_SPLIT_MC_PSC-2-1_spec}
%\end{figure}

\subsection{16 channel prototype}

Based on the availability of a suitable PT adder scheme we designed an integrated phasemeter prototype that was aimed at achieving LISA performance levels at a single frequency with 16 channels. In terms of cross talk we decided to only aim for a suppression level of 60\,dB. This was limited by the cross talk induced on the utilised commercial ADC card (FMC116 from 4DSP \cite{DS_FMC116}). 

\begin{figure}
	\includegraphics[width=\columnwidth]{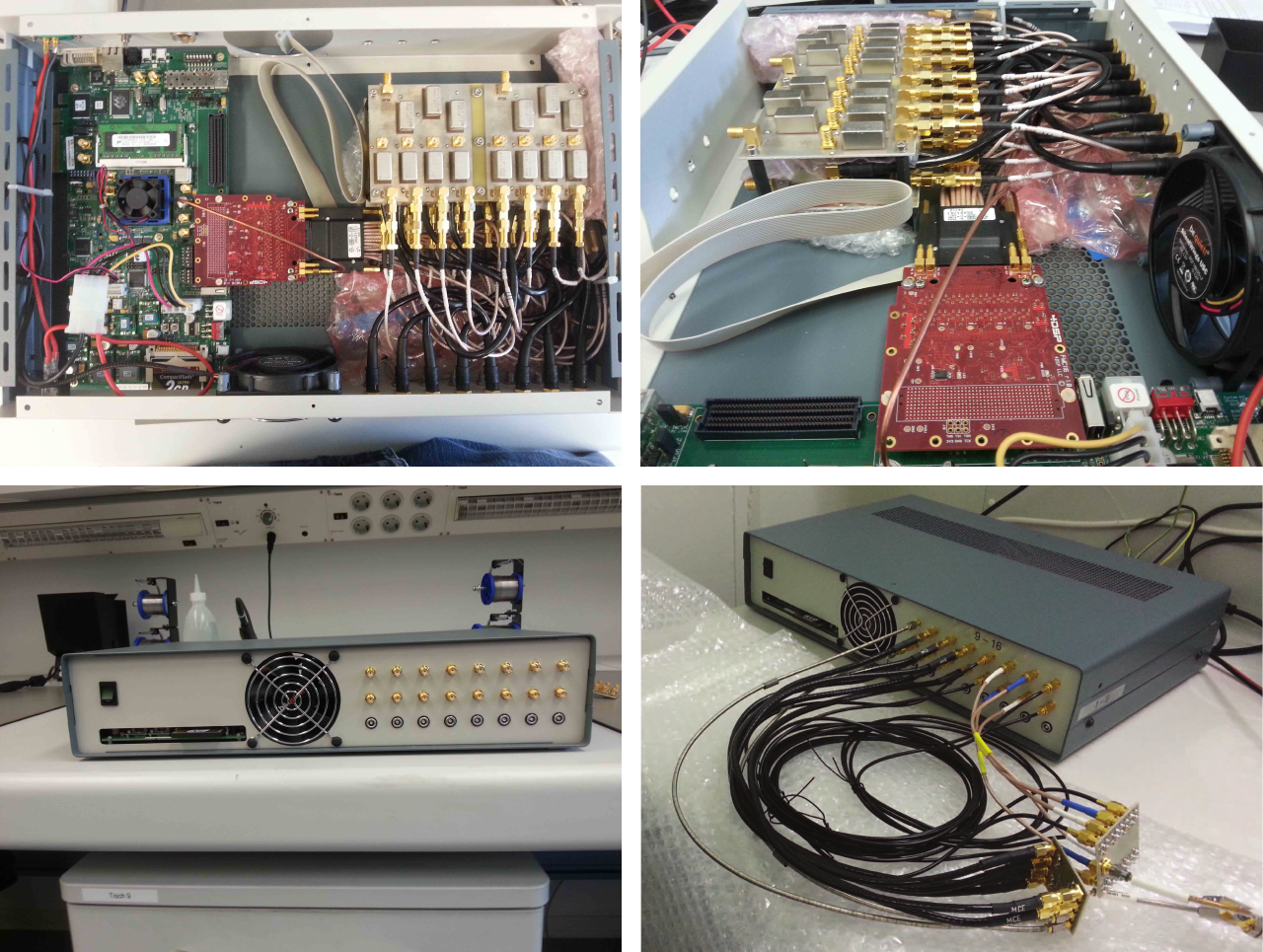}
	\caption{\label{16_pms_photo_low_res}Photographs of the 16 channel phasemeter built using commercially available components in combination with a passive pilot tone distribution and adder circuit. }%
\end{figure}

We chose the implementation relying only on the transformer based power splitter PSC-2-1. Two circuit boards with splitter networks were manufactured and mounted into a common housing with the FPGA board and ADC card, as shown in Figure \ref{16_pms_photo_low_res}. 

\begin{figure}[t]
	\includegraphics[width=\columnwidth]{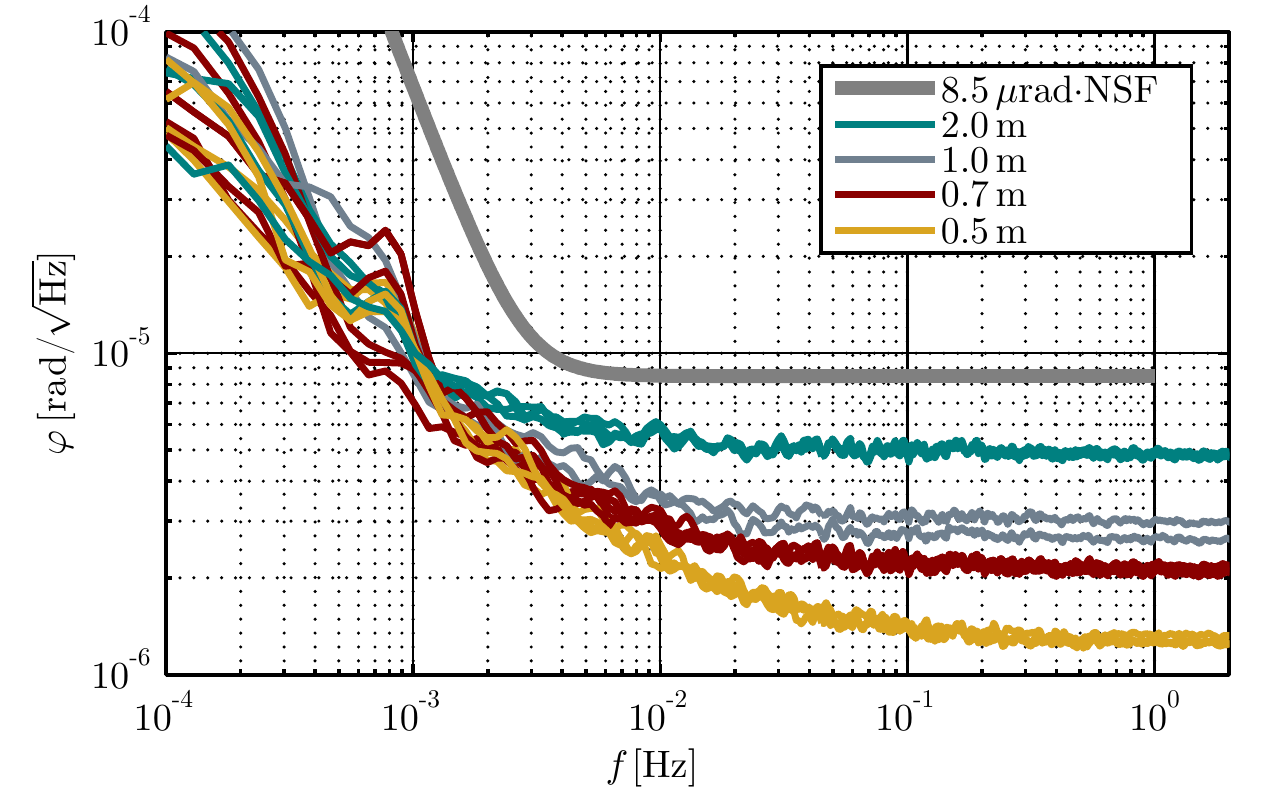}
	\caption{\label{16_pms_cable_spec}Performance measurement using the 16 channel phasemeter. The signal frequency was at 13\,MHz, the pilot tone at 35.3\,MHz. The influence of the different cable lengths (colour coded) on the achieved noise floor is clearly visible. }%
\end{figure}

Using a signal distribution with one two-way resistive splitter, and two eight-way resistive splitters we fed the same input signal into all 16 channels. Again, we were able to achieve $\mu$rad performance with sufficient thermal isolation of the splitters and cables. However, the white noise levels in each channel showed varying levels, as shown in Figure \ref{16_pms_cable_spec}, which were found to correspond to the cable length used for distributing the signals from the splitters to the phasemeter. Using cables with matched lengths allowed us to reduce this noise floor. 
%The dynamics of the input signal, and of the pilot tone, were determined to be a $1/f^{3/2}$ phase noise with an amplitude of about 1\,Hz/$\sqrt{\rm Hz}$ at 1\,Hz. One can easily calculate the effective noise between channels introduced purely due to relative delays. This coupling is linear and proportional to frequency and has a gain at 1\,Hz in the order of $4\pi\tau$, where $\tau$ corresponds to the effective delay or cable length difference divided by the signal propagation speed. This simple model would lead to a cable length dependent $1/\sqrt{f}$ noise floor (due to the signal dynamics). This does not fit our measurement, since the predicted amplitudes are too small and the noise floor is white instead of $1/\sqrt{f}$.

At this point we were not able to fully identify this noise coupling. One speculation is that the white noise is actually caused by a small vector noise present in all channels, which might be caused by impedance mismatches in the signal or PT distribution, by the cross talk between the channels, or a combination of these effects, with the phase of the parasitic tone relative to the actual signal driven by the dynamics of the signal itself. The different cable lengths change the effective coupling in each channel, ultimately leading to the observed white noise floor given by the maximum amplitudes of the small vector noise. A similar effect is observed in the experiment presented in the next section.

 %was So far we did not investigate this noise coupling further, but based on the modest input signal dynamics we can exclude a simple 
%While we were again able to demonstrate the desired phase noise levels, we also found that the measured noise levels were strongly dependent on the used cable lengths, as shown in Figure \ref{16_pms_cable_spec}. 
%
\section{LISA Phasemeter elegant breadboard}
\label{sec:lisapms}
In the scope of the LISA metrology system technology development project we built and tested a full breadboard of the LISA phasemeter \cite{Gerberding2012,Barke2014} (see Figure \ref{phasemeter-full-on-small}). 
%A short overview of the system is given in Figure \ref{phasemeter-full-on-small}. 
For this system we implemented an active AFE using operation amplifiers and a PTD using transformer based power splitters on the clock module and impedance matched distribution on the ADC cards. 

%\begin{figure}
%	\includegraphics[width=\columnwidth]{phasemeter-full-on-small.jpg}
%	\caption{\label{phasemeter-full-on-small} Photograph of the assembled elegant breadboard model of the LISA phasemeter. The system consists of one main board, one clock module, one DAC module and up to five ADC modules \cite{Gerberding2012}.}%
%\end{figure}

\begin{figure}
	\includegraphics[width=\columnwidth]{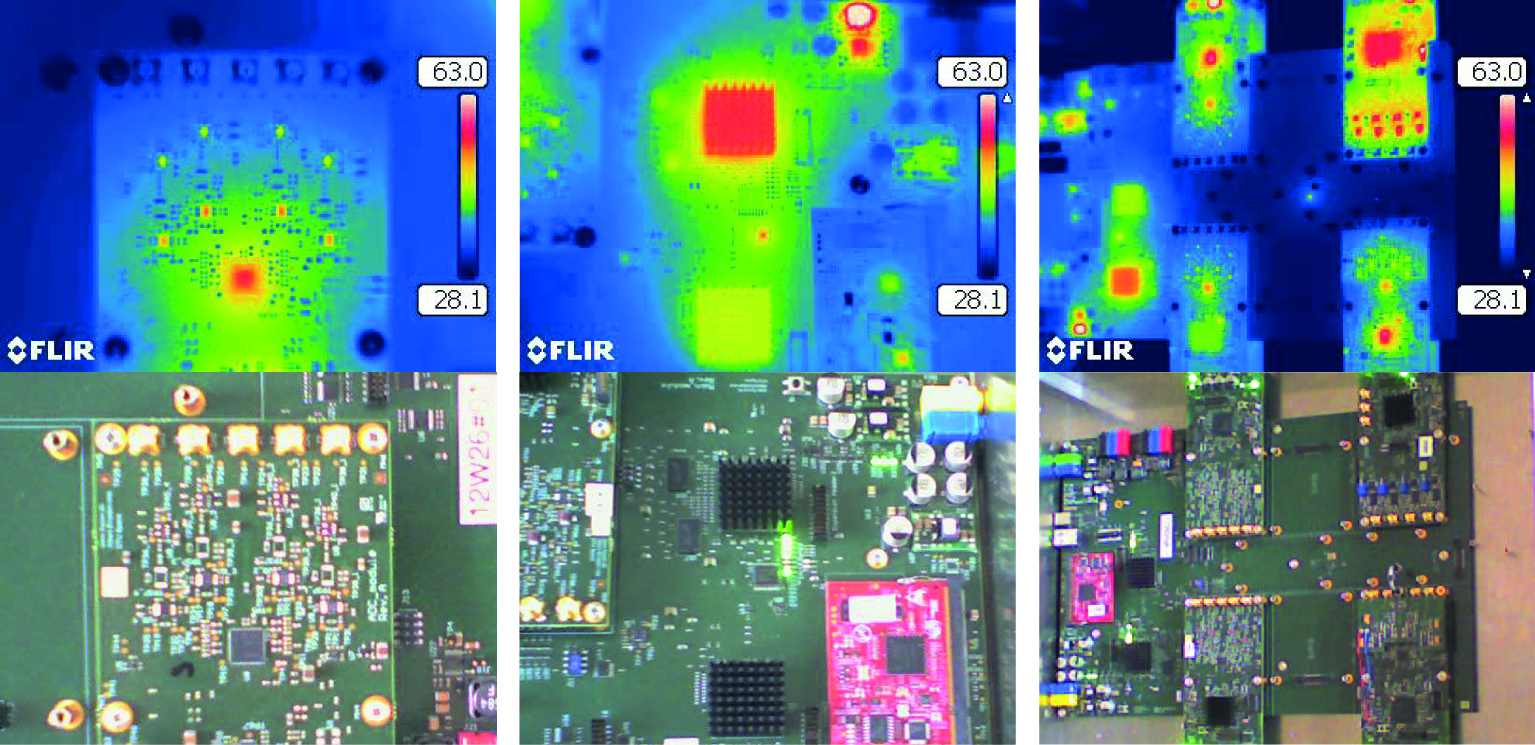}
	\caption{\label{thermal_20mins_low_res} Shown are thermal and corresponding optical pictures of the elegant breadboard model, taken about 20 minutes after start-up in air. The left image shows a close-up of the AFE, showing the heat dissipated by the ADC, which causes a thermal gradient through the AFE. The middle image shows the bridge FPGA and the FFT FPGA \cite{Gerberding2012}, both with passive coolers attached to them, and one of the systems power supplies in the upper right corner. The FFT FPGA has about twice the amount of logic gates and, therefore, a higher power consumption. The right image shows the overall systems with three ADC modules and one DAC module. The DAC card generates significantly more heat. The FPGA temperature also varies strongly on the ADC modules, depending on whether they have a passive cooler mounted or not. }%
\end{figure}

To determine the performance of our system we conducted a comprehensive measurement campaign using LISA like analogue input signals, which were generated by a dedicated signal simulator. These signals included a signal tone that had the dynamics of the expected laser frequency noise (the same conservative spectrum was used in our DSP study \cite{Gerberding2013}) and a constant frequency drift. The synthesised signal also included two smaller side bands at $\pm 1\,$MHz \cite{Heinzel2011}, a weak 1.25\,MHz modulation of pseudo-random noise \cite{Esteban2012,Sutton2013}, and a white noise level with variable signal-to-noise ratio. We performed measurements at various signal frequencies, covering the range from 7\,MHz up to 25\,MHz. The input signal in each case was split using an 8-way resistive splitter and short cables of equal lengths. These were connected to two ADC cards, allowing us to investigate the noise between channels on a single card, and between independent cards.

\begin{figure}
	\includegraphics[width=0.8\columnwidth]{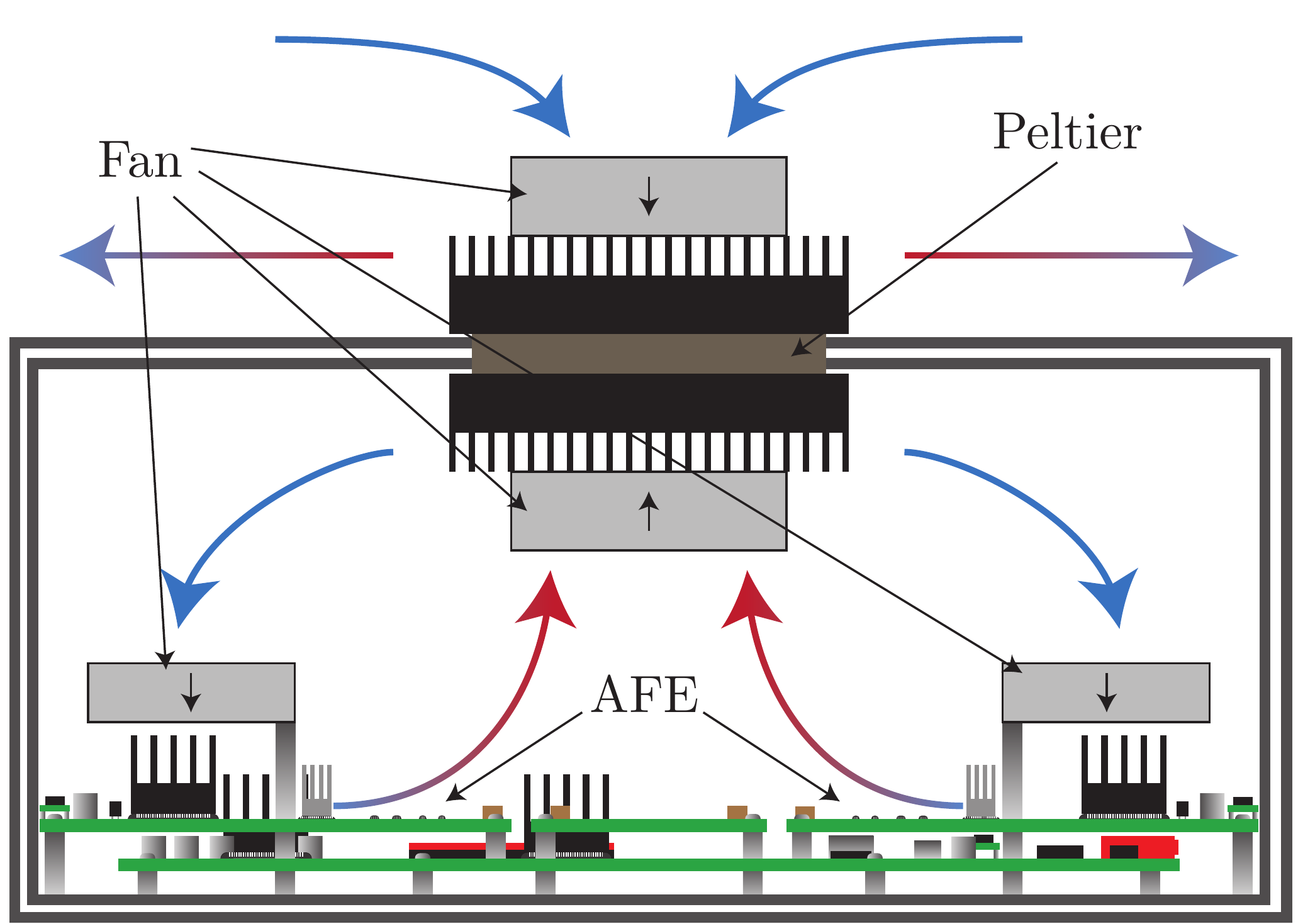}
	\caption{\label{tempstabsketch} Sketch of the active temperature stabilisation scheme. The temperature of the air flowing over the AFE is kept constant by actuating on the Peltier element current. Two large fans ensure the coupling of the inner and out volume to the Peltier and additional fans are monted on each ADC module to optimise the coupling of the airflow to the AFE. Passive heat exchangers are glued onto all ADCs and FPGAs.}%
\end{figure}

\begin{figure}
	\includegraphics[width=\columnwidth]{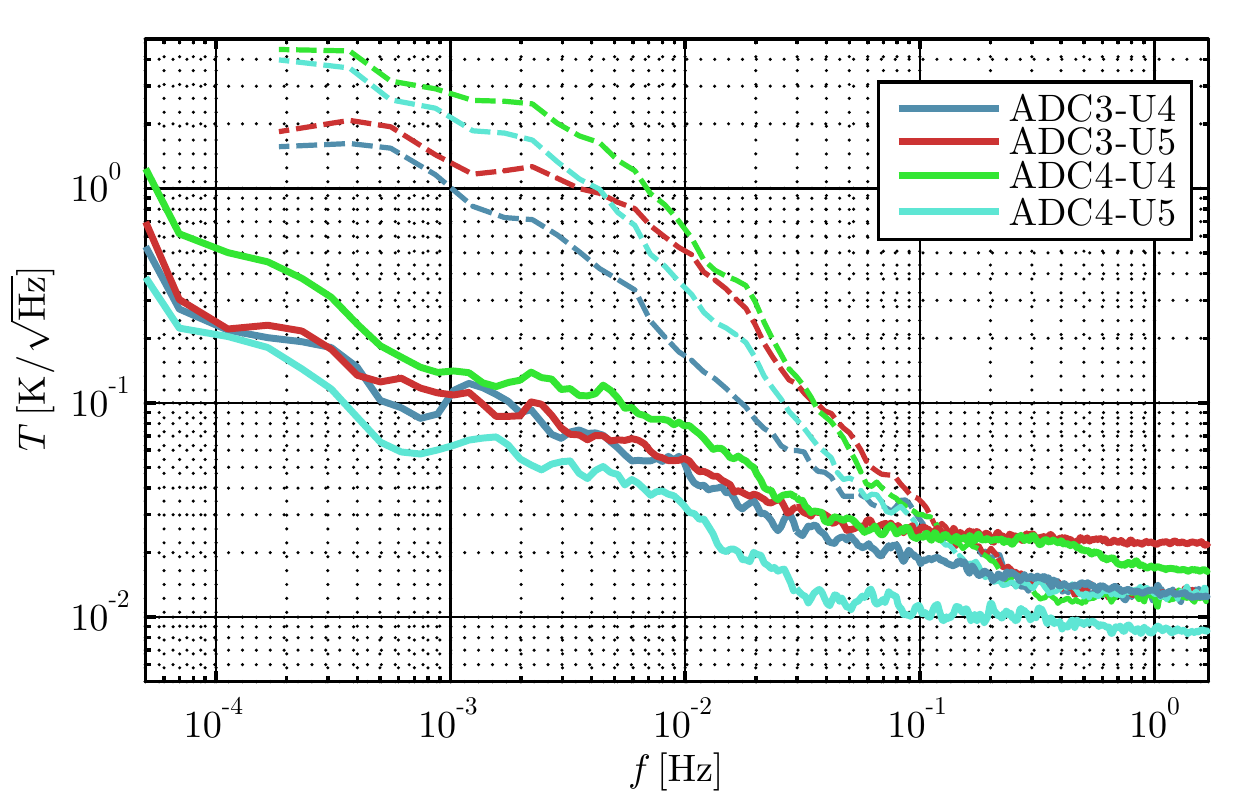}
	\caption{\label{EBB_meas_tempstab} Temperature spectral density measured with the sensors placed next to the AFE of the two ADC cards used during the measurement campaign. Shown are the values while operating the system in air (dashed lines), and while using the active temperature stabilisation in a dedicated housing (solid lines).}%
\end{figure}

The dense placement of the various components on the breadboard caused not only a strong heat dissipation, but also strong temperature fluctuations. An overview of the heat sources is given in Figure \ref{thermal_20mins_low_res}. To achieve the desired performance levels we therefore implemented an active thermal stabilisation of the whole system. To this end, we designed a double-walled housing and integrated a Peltier element into the top, which was coupled to both heat baths via additional fans. A diagram of the stabilisation is shown in Figure \ref{tempstabsketch}. Using a temperature sensor placed close to the AFE, we implemented an active temperature stabilisation which actuated on the current through the Peltier element. Figure \ref{EBB_meas_tempstab} shows the thermal spectra, measured with temperature sensors positioned close to the AFE, on each ADC card with the system open in air and with active stabilisation.
We were able to suppress the temperature fluctuations at the AFE to a level below 200\,mK/$\sqrt{\textrm{Hz}}$ at 1\,mHz. 

\begin{figure}
	\includegraphics[width=\columnwidth]{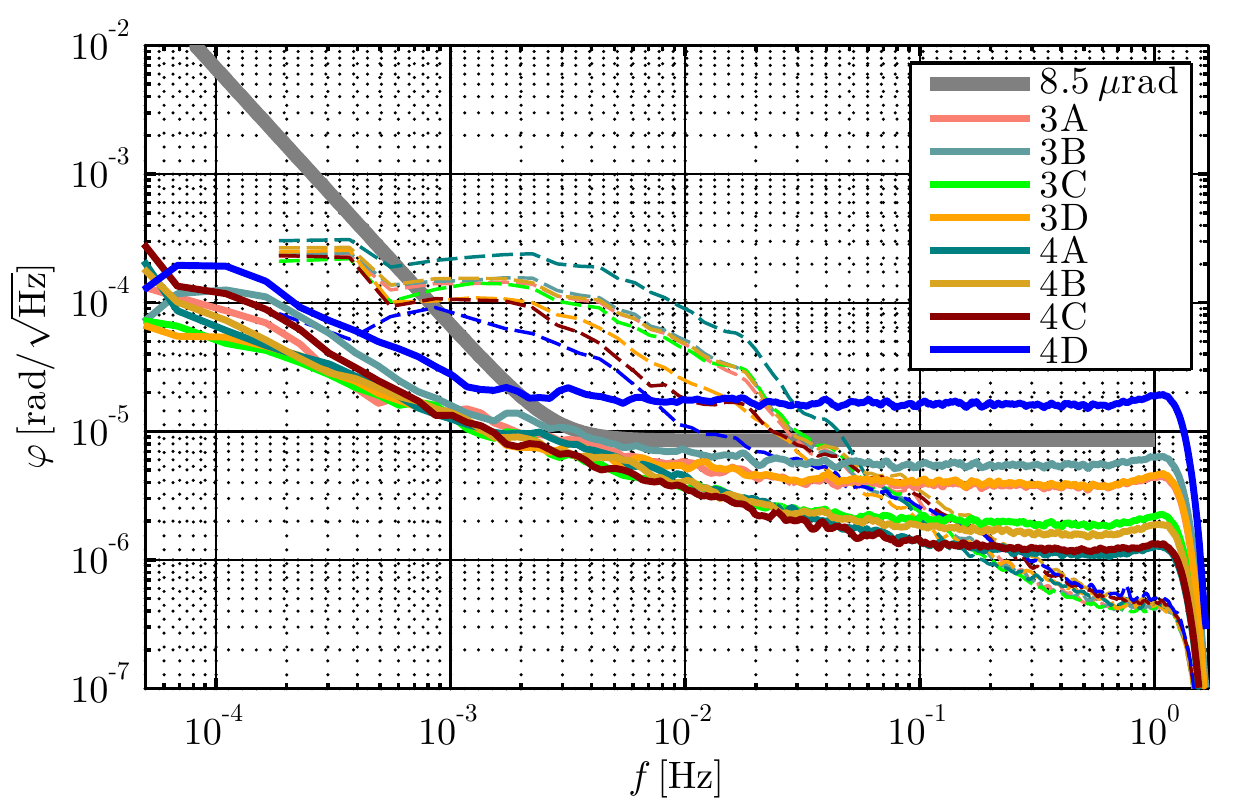}
	\caption{\label{25MHz_comparison} Phase spectral densities for two measurements at 25\,MHz signal frequency, one with temperature stabilisation (solid lines) and one without (dashed lines).}%
\end{figure}

Using that thermal stabilisation we were able to achieve full performance for all frequencies with input white noise corresponding to a SNR of 75\,dBHz and 95\,dBHz; as expected in intersatellite interferometers with long arms and low received optical power. Figure \ref{25MHz_comparison} shows the results of two measurements from our campaign. The first measurement (dashed lines) was done using a standard signal generator, locked to the phasemeter clock, to generate a clean input signal while operating the system in air. The performance is limited in all channels by some 1/f noise (due to thermal noise) and no excess white noise is present in any of the channels. The second measurement (solid lines) was done with the full LISA-like input signal at 75\,dBHz SNR and with the active temperature stabilisation. For most channels we achieved the desired performance and all other measurements using higher SNR and lower signal frequencies showed even better performance. Similar to our earlier investigations with the 16 channel prototype we find, again, a channel dependent white noise floor. From the difference to the first measurement we can directly conclude that this effect is driven by the signal dynamics, which were minimal in the first measurement, leading to no excess white noise. We investigated the cause for observing this coupling without different cable lengths only shortly. We found a dependency of this coupling on the resistor value matching in our resistive power splitter used to distribute the signal. Hence, we speculate that this noise is another example for a small vector noise that is related to the signal dynamics, which is either caused by impedance mismatches, cross talk, or a combination of these effects. 

%\subsection{Results with active thermal stabilisation}

\section{Conclusion}

In this article we have presented a conceptual and experimental analysis of effects in the analogue measurement chain that can spoil the performance of phase readout systems at low frequencies. We have performed various experimental investigations with prototype readout systems and found a wide variety of set-ups that are suitable to achieve performance for certain frequencies. We were able to implement these prototypes with a mixture of commercially available hardware and additionally constructed analogue circuits for distributing and adding a pilot tone.

The design of a full LISA phasemeter prototype, able to operate over the whole desired frequency range, was found to be more demanding, requiring, amongst other things, an active temperature stabilisation of the analogue front-end. With the LISA phasemeter prototype we have demonstrated full phase readout performance of better than $\sqrt{2}\cdot2\pi\,\mu$rad$/\sqrt{\textrm{Hz}}$ for a single channel with LISA like input signals in null-measurement schemes.

Some of our measurements indicate the presence of non-linear noise couplings due to small vector signals. We have discussed the need to take such couplings into account and we described its influence in some detail. Due to the lack of sensitivity to many of these effects, we argue that other types of performance tests have to be implemented in the future, that take such influences into account. Three signal tests involving the whole measurement chain are, in our perspective, the preferred method to do that.

\begin{acknowledgments}
We acknowledge funding by the European Space Agency (ESA) within the technology development project ``LISA Metrology System'' and ``Optical Bench Development for LISA''. The authors also gratefully acknowledge support by the International Max-Planck Research School for Gravitational Wave Astronomy (IMPRS-GW), by the Centre for Quantum Engineering and Space-Time Research (QUEST) and by Deutsches Zentrum f\"ur Luft- und Raumfahrt (DLR) with funding from the Bundesministerium für Wirtschaft und Technologie (project reference 50 OQ 0601). 
\end{acknowledgments}

%\bibliography{../../../../literature/PhD_AEI}

\begin{thebibliography}{25}%
	\makeatletter
	\providecommand \@ifxundefined [1]{%
		\@ifx{#1\undefined}
	}%
	\providecommand \@ifnum [1]{%
		\ifnum #1\expandafter \@firstoftwo
		\else \expandafter \@secondoftwo
		\fi
	}%
	\providecommand \@ifx [1]{%
		\ifx #1\expandafter \@firstoftwo
		\else \expandafter \@secondoftwo
		\fi
	}%
	\providecommand \natexlab [1]{#1}%
	\providecommand \enquote  [1]{``#1''}%
	\providecommand \bibnamefont  [1]{#1}%
	\providecommand \bibfnamefont [1]{#1}%
	\providecommand \citenamefont [1]{#1}%
	\providecommand \href@noop [0]{\@secondoftwo}%
	\providecommand \href [0]{\begingroup \@sanitize@url \@href}%
	\providecommand \@href[1]{\@@startlink{#1}\@@href}%
	\providecommand \@@href[1]{\endgroup#1\@@endlink}%
	\providecommand \@sanitize@url [0]{\catcode `\\12\catcode `\$12\catcode
		`\&12\catcode `\#12\catcode `\^12\catcode `\_12\catcode `\%12\relax}%
	\providecommand \@@startlink[1]{}%
	\providecommand \@@endlink[0]{}%
	\providecommand \url  [0]{\begingroup\@sanitize@url \@url }%
	\providecommand \@url [1]{\endgroup\@href {#1}{\urlprefix }}%
	\providecommand \urlprefix  [0]{URL }%
	\providecommand \Eprint [0]{\href }%
	\providecommand \doibase [0]{http://dx.doi.org/}%
	\providecommand \selectlanguage [0]{\@gobble}%
	\providecommand \bibinfo  [0]{\@secondoftwo}%
	\providecommand \bibfield  [0]{\@secondoftwo}%
	\providecommand \translation [1]{[#1]}%
	\providecommand \BibitemOpen [0]{}%
	\providecommand \bibitemStop [0]{}%
	\providecommand \bibitemNoStop [0]{.\EOS\space}%
	\providecommand \EOS [0]{\spacefactor3000\relax}%
	\providecommand \BibitemShut  [1]{\csname bibitem#1\endcsname}%
	\let\auto@bib@innerbib\@empty
	%</preamble>
	\bibitem [{\citenamefont {Danzmann}\ \emph {et~al.}(2013)\citenamefont
		{Danzmann} \emph {et~al.}}]{Danzmann2013}%
	\BibitemOpen
	\bibfield  {author} {\bibinfo {author} {\bibfnamefont {K.}~\bibnamefont
			{Danzmann}} \emph {et~al.},\ }\href {http://arxiv.org/abs/1305.5720}
	{\enquote {\bibinfo {title} {{The Gravitational Universe: Whitepaper for the
					ESA L2/L3 selection}},}\ }\bibinfo {howpublished}
	{\href{http://arxiv.org/abs/1305.5720}{arXiv preprint arXiv:1305.5720}}
	(\bibinfo {year} {2013})\BibitemShut {NoStop}%
	\bibitem [{\citenamefont {Sheard}\ \emph {et~al.}(2012)\citenamefont {Sheard},
		\citenamefont {Heinzel}, \citenamefont {Danzmann}, \citenamefont {Shaddock},
		\citenamefont {Klipstein},\ and\ \citenamefont {Folkner}}]{Sheard2012}%
	\BibitemOpen
	\bibfield  {author} {\bibinfo {author} {\bibfnamefont {B.~S.}\ \bibnamefont
			{Sheard}}, \bibinfo {author} {\bibfnamefont {G.}~\bibnamefont {Heinzel}},
		\bibinfo {author} {\bibfnamefont {K.}~\bibnamefont {Danzmann}}, \bibinfo
		{author} {\bibfnamefont {D.~A.}\ \bibnamefont {Shaddock}}, \bibinfo {author}
		{\bibfnamefont {W.~M.}\ \bibnamefont {Klipstein}}, \ and\ \bibinfo {author}
		{\bibfnamefont {W.~M.}\ \bibnamefont {Folkner}},\ }\href@noop {} {\bibfield
		{journal} {\bibinfo  {journal} {Journal of Geodesy}\ }\textbf {\bibinfo
			{volume} {86}},\ \bibinfo {pages} {1083} (\bibinfo {year}
		{2012})}\BibitemShut {NoStop}%
	\bibitem [{\citenamefont {Shaddock}\ \emph {et~al.}(2006)\citenamefont
		{Shaddock}, \citenamefont {Ware}, \citenamefont {Halverson}, \citenamefont
		{Spero},\ and\ \citenamefont {Klipstein}}]{Shaddock2006}%
	\BibitemOpen
	\bibfield  {author} {\bibinfo {author} {\bibfnamefont {D.}~\bibnamefont
			{Shaddock}}, \bibinfo {author} {\bibfnamefont {B.}~\bibnamefont {Ware}},
		\bibinfo {author} {\bibfnamefont {P.}~\bibnamefont {Halverson}}, \bibinfo
		{author} {\bibfnamefont {R.~E.}\ \bibnamefont {Spero}}, \ and\ \bibinfo
		{author} {\bibfnamefont {B.}~\bibnamefont {Klipstein}},\ }\href@noop {}
	{\bibfield  {journal} {\bibinfo  {journal} {AIP Conf. Proc.}\ }\textbf
		{\bibinfo {volume} {{873}}},\ \bibinfo {pages} {689} (\bibinfo {year}
		{2006})}\BibitemShut {NoStop}%
	\bibitem [{\citenamefont {Gerberding}\ \emph {et~al.}(2013)\citenamefont
		{Gerberding}, \citenamefont {Sheard}, \citenamefont {Bykov}, \citenamefont
		{Kullmann}, \citenamefont {Delgado}, \citenamefont {Danzmann},\ and\
		\citenamefont {Heinzel}}]{Gerberding2013}%
	\BibitemOpen
	\bibfield  {author} {\bibinfo {author} {\bibfnamefont {O.}~\bibnamefont
			{Gerberding}}, \bibinfo {author} {\bibfnamefont {B.}~\bibnamefont {Sheard}},
		\bibinfo {author} {\bibfnamefont {I.}~\bibnamefont {Bykov}}, \bibinfo
		{author} {\bibfnamefont {J.}~\bibnamefont {Kullmann}}, \bibinfo {author}
		{\bibfnamefont {J.~J.~E.}\ \bibnamefont {Delgado}}, \bibinfo {author}
		{\bibfnamefont {K.}~\bibnamefont {Danzmann}}, \ and\ \bibinfo {author}
		{\bibfnamefont {G.}~\bibnamefont {Heinzel}},\ }\href@noop {} {\bibfield
		{journal} {\bibinfo  {journal} {Classical and Quantum Gravity}\ }\textbf
		{\bibinfo {volume} {30}},\ \bibinfo {pages} {235029} (\bibinfo {year}
		{2013})}\BibitemShut {NoStop}%
	\bibitem [{\citenamefont {Francis}\ \emph {et~al.}(2014)\citenamefont
		{Francis}, \citenamefont {Lam}, \citenamefont {McKenzie}, \citenamefont
		{Sutton}, \citenamefont {Ward}, \citenamefont {McClelland},\ and\
		\citenamefont {Shaddock}}]{Francis2014}%
	\BibitemOpen
	\bibfield  {author} {\bibinfo {author} {\bibfnamefont {S.~P.}\ \bibnamefont
			{Francis}}, \bibinfo {author} {\bibfnamefont {T.~T.-Y.}\ \bibnamefont {Lam}},
		\bibinfo {author} {\bibfnamefont {K.}~\bibnamefont {McKenzie}}, \bibinfo
		{author} {\bibfnamefont {A.~J.}\ \bibnamefont {Sutton}}, \bibinfo {author}
		{\bibfnamefont {R.~L.}\ \bibnamefont {Ward}}, \bibinfo {author}
		{\bibfnamefont {D.~E.}\ \bibnamefont {McClelland}}, \ and\ \bibinfo {author}
		{\bibfnamefont {D.~A.}\ \bibnamefont {Shaddock}},\ }\href {\doibase
		10.1364/OL.39.005251} {\bibfield  {journal} {\bibinfo  {journal} {Opt.
				Lett.}\ }\textbf {\bibinfo {volume} {39}},\ \bibinfo {pages} {5251} (\bibinfo
		{year} {2014})}\BibitemShut {NoStop}%
	\bibitem [{\citenamefont {Yu}, \citenamefont {Mitryk},\ and\ \citenamefont
		{Mueller}(2014)}]{Yu2014}%
	\BibitemOpen
	\bibfield  {author} {\bibinfo {author} {\bibfnamefont {Y.}~\bibnamefont
			{Yu}}, \bibinfo {author} {\bibfnamefont {S.}~\bibnamefont {Mitryk}}, \ and\
		\bibinfo {author} {\bibfnamefont {G.}~\bibnamefont {Mueller}},\ }\href
	{\doibase 10.1103/PhysRevD.90.062005} {\bibfield  {journal} {\bibinfo
			{journal} {Phys. Rev. D}\ }\textbf {\bibinfo {volume} {90}},\ \bibinfo
		{pages} {062005} (\bibinfo {year} {2014})}\BibitemShut {NoStop}%
	\bibitem [{\citenamefont {Liu}\ \emph {et~al.}(2014)\citenamefont {Liu},
		\citenamefont {Dong}, \citenamefont {Li}, \citenamefont {Luo},\ and\
		\citenamefont {Jin}}]{Liu2014}%
	\BibitemOpen
	\bibfield  {author} {\bibinfo {author} {\bibfnamefont {H.-S.}\ \bibnamefont
			{Liu}}, \bibinfo {author} {\bibfnamefont {Y.-H.}\ \bibnamefont {Dong}},
		\bibinfo {author} {\bibfnamefont {Y.-Q.}\ \bibnamefont {Li}}, \bibinfo
		{author} {\bibfnamefont {Z.-R.}\ \bibnamefont {Luo}}, \ and\ \bibinfo
		{author} {\bibfnamefont {G.}~\bibnamefont {Jin}},\ }\href {\doibase
		10.1063/1.4865121} {\bibfield  {journal} {\bibinfo  {journal} {Review of
				Scientific Instruments}\ }\textbf {\bibinfo {volume} {85}},\ \bibinfo {pages}
		{024503} (\bibinfo {year} {2014})}\BibitemShut {NoStop}%
	\bibitem [{\citenamefont {Liang}\ \emph {et~al.}(2015)\citenamefont {Liang},
		\citenamefont {Duan}, \citenamefont {Xiao}, \citenamefont {Wei},\ and\
		\citenamefont {Yeh}}]{Liang2015}%
	\BibitemOpen
	\bibfield  {author} {\bibinfo {author} {\bibfnamefont {Y.-R.}\ \bibnamefont
			{Liang}}, \bibinfo {author} {\bibfnamefont {H.-Z.}\ \bibnamefont {Duan}},
		\bibinfo {author} {\bibfnamefont {X.-L.}\ \bibnamefont {Xiao}}, \bibinfo
		{author} {\bibfnamefont {B.-B.}\ \bibnamefont {Wei}}, \ and\ \bibinfo
		{author} {\bibfnamefont {H.-C.}\ \bibnamefont {Yeh}},\ }\href {\doibase
		http://dx.doi.org/10.1063/1.4905579} {\bibfield  {journal} {\bibinfo
			{journal} {Review of Scientific Instruments}\ }\textbf {\bibinfo {volume}
			{86}},\ \bibinfo {eid} {016106} (\bibinfo {year} {2015})}\BibitemShut
	{NoStop}%
	\bibitem [{\citenamefont {Gerberding}\ \emph {et~al.}(2012)\citenamefont
		{Gerberding}, \citenamefont {Barke}, \citenamefont {Bykov}, \citenamefont
		{Danzmann}, \citenamefont {Enddaard}, \citenamefont {Esteban}, \citenamefont
		{Gianolio}, \citenamefont {Hansen}, \citenamefont {Heinzel}, \citenamefont
		{Hornstrup}, \citenamefont {Jennrich}, \citenamefont {Kullmann},
		\citenamefont {Pedersen}, \citenamefont {Rasmussen}, \citenamefont {Sodnik},\
		and\ \citenamefont {Suess}}]{Gerberding2012}%
	\BibitemOpen
	\bibfield  {author} {\bibinfo {author} {\bibfnamefont {O.}~\bibnamefont
			{Gerberding}}, \bibinfo {author} {\bibfnamefont {S.}~\bibnamefont {Barke}},
		\bibinfo {author} {\bibfnamefont {I.}~\bibnamefont {Bykov}}, \bibinfo
		{author} {\bibfnamefont {K.}~\bibnamefont {Danzmann}}, \bibinfo {author}
		{\bibfnamefont {A.}~\bibnamefont {Enddaard}}, \bibinfo {author}
		{\bibfnamefont {J.~J.}\ \bibnamefont {Esteban}}, \bibinfo {author}
		{\bibfnamefont {A.}~\bibnamefont {Gianolio}}, \bibinfo {author}
		{\bibfnamefont {T.~V.}\ \bibnamefont {Hansen}}, \bibinfo {author}
		{\bibfnamefont {G.}~\bibnamefont {Heinzel}}, \bibinfo {author} {\bibfnamefont
			{A.}~\bibnamefont {Hornstrup}}, \bibinfo {author} {\bibfnamefont
			{O.}~\bibnamefont {Jennrich}}, \bibinfo {author} {\bibfnamefont
			{J.}~\bibnamefont {Kullmann}}, \bibinfo {author} {\bibfnamefont {S.~M.}\
			\bibnamefont {Pedersen}}, \bibinfo {author} {\bibfnamefont {T.}~\bibnamefont
			{Rasmussen}}, \bibinfo {author} {\bibfnamefont {Z.}~\bibnamefont {Sodnik}}, \
		and\ \bibinfo {author} {\bibfnamefont {M.}~\bibnamefont {Suess}},\ }in\
	\href@noop {} {\emph {\bibinfo {booktitle} {ASP Conference Series; 9th LISA
				Symposium}}},\ \bibinfo {series} {ASP Conference Series}, Vol.\ \bibinfo
	{volume} {467},\ \bibinfo {editor} {edited by\ \bibinfo {editor}
		{\bibfnamefont {G.}~\bibnamefont {Auger}}, \bibinfo {editor} {\bibfnamefont
			{P.}~\bibnamefont {Binetruy}}, \ and\ \bibinfo {editor} {\bibfnamefont
			{E.}~\bibnamefont {Plagnol}}}\ (\bibinfo {year} {2012})\ pp.\ \bibinfo
	{pages} {271 -- 275}\BibitemShut {NoStop}%
	\bibitem [{\citenamefont {Barke}\ \emph {et~al.}(2014)\citenamefont {Barke},
		\citenamefont {Brause}, \citenamefont {Bykov}, \citenamefont
		{Esteban~Delgado}, \citenamefont {Enggaard}, \citenamefont {Gerberding},
		\citenamefont {Heinzel}, \citenamefont {Kullmann}, \citenamefont {Pedersen},\
		and\ \citenamefont {Rasmussen}}]{Barke2014}%
	\BibitemOpen
	\bibfield  {author} {\bibinfo {author} {\bibfnamefont {S.}~\bibnamefont
			{Barke}}, \bibinfo {author} {\bibfnamefont {N.}~\bibnamefont {Brause}},
		\bibinfo {author} {\bibfnamefont {I.}~\bibnamefont {Bykov}}, \bibinfo
		{author} {\bibfnamefont {J.~J.}\ \bibnamefont {Esteban~Delgado}}, \bibinfo
		{author} {\bibfnamefont {A.}~\bibnamefont {Enggaard}}, \bibinfo {author}
		{\bibfnamefont {O.}~\bibnamefont {Gerberding}}, \bibinfo {author}
		{\bibfnamefont {G.}~\bibnamefont {Heinzel}}, \bibinfo {author} {\bibfnamefont
			{J.}~\bibnamefont {Kullmann}}, \bibinfo {author} {\bibfnamefont {S.~M.}\
			\bibnamefont {Pedersen}}, \ and\ \bibinfo {author} {\bibfnamefont
			{T.}~\bibnamefont {Rasmussen}},\ }\href
	{http://hdl.handle.net/11858/00-001M-0000-0023-E266-6} {\enquote {\bibinfo
			{title} {{LISA} metrology system-final report},}\ }\bibinfo {howpublished}
	{online: http://hdl.handle.net/11858/00-001M-0000-0023-E266-6} (\bibinfo
	{year} {2014})\BibitemShut {NoStop}%
	\bibitem [{\citenamefont {Tr\"{o}bs}\ \emph {et~al.}(2012)\citenamefont
		{Tr\"{o}bs}, \citenamefont {dArcio}, \citenamefont {Barke}, \citenamefont
		{Bogenstahl}, \citenamefont {Bykov}, \citenamefont {Dehne}, \citenamefont
		{Diekmann}, \citenamefont {Fitzsimons}, \citenamefont {Fleddermann},
		\citenamefont {Gerberding} \emph {et~al.}}]{Trobs2012}%
	\BibitemOpen
	\bibfield  {author} {\bibinfo {author} {\bibfnamefont {M.}~\bibnamefont
			{Tr\"{o}bs}}, \bibinfo {author} {\bibfnamefont {L.}~\bibnamefont {dArcio}},
		\bibinfo {author} {\bibfnamefont {S.}~\bibnamefont {Barke}}, \bibinfo
		{author} {\bibfnamefont {J.}~\bibnamefont {Bogenstahl}}, \bibinfo {author}
		{\bibfnamefont {I.}~\bibnamefont {Bykov}}, \bibinfo {author} {\bibfnamefont
			{M.}~\bibnamefont {Dehne}}, \bibinfo {author} {\bibfnamefont
			{C.}~\bibnamefont {Diekmann}}, \bibinfo {author} {\bibfnamefont
			{E.}~\bibnamefont {Fitzsimons}}, \bibinfo {author} {\bibfnamefont
			{R.}~\bibnamefont {Fleddermann}}, \bibinfo {author} {\bibfnamefont
			{O.}~\bibnamefont {Gerberding}},  \emph {et~al.},\ }in\ \href@noop {} {\emph
		{\bibinfo {booktitle} {Proceedings of the 9th International Conference on
				Space Optics, Ajaccio}}}\ (\bibinfo {year} {2012})\BibitemShut {NoStop}%
	\bibitem [{\citenamefont {Spero}\ \emph {et~al.}(2011)\citenamefont {Spero},
		\citenamefont {Bachman}, \citenamefont {de~Vine}, \citenamefont {Dickson},
		\citenamefont {Klipstein}, \citenamefont {Ozawa}, \citenamefont {McKenzie},
		\citenamefont {Shaddock}, \citenamefont {Robison}, \citenamefont {Sutton},\
		and\ \citenamefont {Ware}}]{Spero2011}%
	\BibitemOpen
	\bibfield  {author} {\bibinfo {author} {\bibfnamefont {R.}~\bibnamefont
			{Spero}}, \bibinfo {author} {\bibfnamefont {B.}~\bibnamefont {Bachman}},
		\bibinfo {author} {\bibfnamefont {G.}~\bibnamefont {de~Vine}}, \bibinfo
		{author} {\bibfnamefont {J.}~\bibnamefont {Dickson}}, \bibinfo {author}
		{\bibfnamefont {W.~M.}\ \bibnamefont {Klipstein}}, \bibinfo {author}
		{\bibfnamefont {T.}~\bibnamefont {Ozawa}}, \bibinfo {author} {\bibfnamefont
			{K.}~\bibnamefont {McKenzie}}, \bibinfo {author} {\bibfnamefont {D.~A.}\
			\bibnamefont {Shaddock}}, \bibinfo {author} {\bibfnamefont {D.}~\bibnamefont
			{Robison}}, \bibinfo {author} {\bibfnamefont {A.}~\bibnamefont {Sutton}}, \
		and\ \bibinfo {author} {\bibfnamefont {B.}~\bibnamefont {Ware}},\ }\href@noop
	{} {\bibfield  {journal} {\bibinfo  {journal} {Class. Quantum Grav.}\
		}\textbf {\bibinfo {volume} {28}},\ \bibinfo {pages} {094007} (\bibinfo
		{year} {2011})}\BibitemShut {NoStop}%
	\bibitem [{\citenamefont {Sch\"{u}tze}\ \emph {et~al.}(2014)\citenamefont
		{Sch\"{u}tze}, \citenamefont {Stede}, \citenamefont {M\"{u}ller},
		\citenamefont {Gerberding}, \citenamefont {Bandikova}, \citenamefont
		{Sheard}, \citenamefont {Heinzel},\ and\ \citenamefont
		{Danzmann}}]{Schutze2014}%
	\BibitemOpen
	\bibfield  {author} {\bibinfo {author} {\bibfnamefont {D.}~\bibnamefont
			{Sch\"{u}tze}}, \bibinfo {author} {\bibfnamefont {G.}~\bibnamefont {Stede}},
		\bibinfo {author} {\bibfnamefont {V.}~\bibnamefont {M\"{u}ller}}, \bibinfo
		{author} {\bibfnamefont {O.}~\bibnamefont {Gerberding}}, \bibinfo {author}
		{\bibfnamefont {T.}~\bibnamefont {Bandikova}}, \bibinfo {author}
		{\bibfnamefont {B.~S.}\ \bibnamefont {Sheard}}, \bibinfo {author}
		{\bibfnamefont {G.}~\bibnamefont {Heinzel}}, \ and\ \bibinfo {author}
		{\bibfnamefont {K.}~\bibnamefont {Danzmann}},\ }\href {\doibase
		10.1364/OE.22.024117} {\bibfield  {journal} {\bibinfo  {journal} {Opt.
				Express}\ }\textbf {\bibinfo {volume} {22}},\ \bibinfo {pages} {24117}
		(\bibinfo {year} {2014})}\BibitemShut {NoStop}%
	\bibitem [{\citenamefont {Cervantes}\ \emph {et~al.}(2011)\citenamefont
		{Cervantes}, \citenamefont {Livas}, \citenamefont {Silverberg}, \citenamefont
		{Buchanan},\ and\ \citenamefont {Stebbins}}]{Cervantes2011}%
	\BibitemOpen
	\bibfield  {author} {\bibinfo {author} {\bibfnamefont {F.~G.}\ \bibnamefont
			{Cervantes}}, \bibinfo {author} {\bibfnamefont {J.}~\bibnamefont {Livas}},
		\bibinfo {author} {\bibfnamefont {R.}~\bibnamefont {Silverberg}}, \bibinfo
		{author} {\bibfnamefont {E.}~\bibnamefont {Buchanan}}, \ and\ \bibinfo
		{author} {\bibfnamefont {R.}~\bibnamefont {Stebbins}},\ }\href
	{http://stacks.iop.org/0264-9381/28/i=9/a=094010} {\bibfield  {journal}
		{\bibinfo  {journal} {Classical and Quantum Gravity}\ }\textbf {\bibinfo
			{volume} {28}},\ \bibinfo {pages} {094010} (\bibinfo {year}
		{2011})}\BibitemShut {NoStop}%
	\bibitem [{\citenamefont {Wannamaker}(2003)}]{Wannamaker2003}%
	\BibitemOpen
	\bibfield  {author} {\bibinfo {author} {\bibfnamefont {R.~A.}\ \bibnamefont
			{Wannamaker}},\ }\href@noop {} {\emph {\bibinfo {title} {The theory of
				dithered quantization.}}}\ (\bibinfo  {publisher} {University of Waterloo},\
	\bibinfo {year} {2003})\BibitemShut {NoStop}%
	\bibitem [{\citenamefont {Carter}(2003)}]{Carter2003}%
	\BibitemOpen
	\bibfield  {author} {\bibinfo {author} {\bibfnamefont {B.}~\bibnamefont
			{Carter}},\ }\href@noop {} {\emph {\bibinfo {title} {Op Amps for everyone}}}\
	(\bibinfo  {publisher} {Elsevier},\ \bibinfo {year} {2003})\BibitemShut
	{NoStop}%
	\bibitem [{\citenamefont {Boudot}\ and\ \citenamefont
		{Rubiola}(2012)}]{Boudot2012}%
	\BibitemOpen
	\bibfield  {author} {\bibinfo {author} {\bibfnamefont {R.}~\bibnamefont
			{Boudot}}\ and\ \bibinfo {author} {\bibfnamefont {E.}~\bibnamefont
			{Rubiola}},\ }\href@noop {} {\bibfield  {journal} {\bibinfo  {journal}
			{Ultrasonics, Ferroelectrics and Frequency Control, IEEE Transactions on}\
		}\textbf {\bibinfo {volume} {59}},\ \bibinfo {pages} {2613} (\bibinfo {year}
		{2012})}\BibitemShut {NoStop}%
	\bibitem [{\citenamefont {Hechenblaikner}(2013)}]{Hechenblaikner2013}%
	\BibitemOpen
	\bibfield  {author} {\bibinfo {author} {\bibfnamefont {G.}~\bibnamefont
			{Hechenblaikner}},\ }\href {\doibase 10.1364/JOSAA.30.000941} {\bibfield
		{journal} {\bibinfo  {journal} {J. Opt. Soc. Am. A}\ }\textbf {\bibinfo
			{volume} {30}},\ \bibinfo {pages} {941} (\bibinfo {year} {2013})}\BibitemShut
	{NoStop}%
	\bibitem [{\citenamefont {Tr{\"o}bs}\ \emph {et~al.}(2013)\citenamefont
		{Tr{\"o}bs}, \citenamefont {d'Arcio}, \citenamefont {Barke}, \citenamefont
		{Bogenstahl}, \citenamefont {Diekmann}, \citenamefont {Fitzsimons},
		\citenamefont {Gerberding}, \citenamefont {Hennig}, \citenamefont {Hey},
		\citenamefont {Hogenhuis} \emph {et~al.}}]{Trobs2013}%
	\BibitemOpen
	\bibfield  {author} {\bibinfo {author} {\bibfnamefont {M.}~\bibnamefont
			{Tr{\"o}bs}}, \bibinfo {author} {\bibfnamefont {L.}~\bibnamefont {d'Arcio}},
		\bibinfo {author} {\bibfnamefont {S.}~\bibnamefont {Barke}}, \bibinfo
		{author} {\bibfnamefont {J.}~\bibnamefont {Bogenstahl}}, \bibinfo {author}
		{\bibfnamefont {C.}~\bibnamefont {Diekmann}}, \bibinfo {author}
		{\bibfnamefont {E.}~\bibnamefont {Fitzsimons}}, \bibinfo {author}
		{\bibfnamefont {O.}~\bibnamefont {Gerberding}}, \bibinfo {author}
		{\bibfnamefont {J.}~\bibnamefont {Hennig}}, \bibinfo {author} {\bibfnamefont
			{F.}~\bibnamefont {Hey}}, \bibinfo {author} {\bibfnamefont {H.}~\bibnamefont
			{Hogenhuis}},  \emph {et~al.},\ }in\ \href@noop {} {\emph {\bibinfo
			{booktitle} {Astronomical Society of the Pacific Conference Series}}},\ Vol.\
	\bibinfo {volume} {467}\ (\bibinfo {year} {2013})\ p.\ \bibinfo {pages}
	{233}\BibitemShut {NoStop}%
	\bibitem [{\citenamefont {Joshi}\ \emph {et~al.}(2012)\citenamefont {Joshi},
		\citenamefont {Datta}, \citenamefont {Rue}, \citenamefont {Livas},
		\citenamefont {Silverberg},\ and\ \citenamefont
		{Guzman~Cervantes}}]{Joshi2012}%
	\BibitemOpen
	\bibfield  {author} {\bibinfo {author} {\bibfnamefont {A.}~\bibnamefont
			{Joshi}}, \bibinfo {author} {\bibfnamefont {S.}~\bibnamefont {Datta}},
		\bibinfo {author} {\bibfnamefont {J.}~\bibnamefont {Rue}}, \bibinfo {author}
		{\bibfnamefont {J.}~\bibnamefont {Livas}}, \bibinfo {author} {\bibfnamefont
			{R.}~\bibnamefont {Silverberg}}, \ and\ \bibinfo {author} {\bibfnamefont
			{F.}~\bibnamefont {Guzman~Cervantes}},\ }in\ \href {\doibase
		10.1117/12.918285} {\emph {\bibinfo {booktitle} {SPIE Astronomical
				Telescopes+ Instrumentation}}},\ Vol.\ \bibinfo {volume} {8453}\ (\bibinfo
	{year} {2012})\ pp.\ \bibinfo {pages} {84532G--84532G--10}\BibitemShut
	{NoStop}%
	\bibitem [{\citenamefont {{4DSP}}(2011{\natexlab{a}})}]{DS_FMC107}%
	\BibitemOpen
	\bibfield  {author} {\bibinfo {author} {\bibnamefont {{4DSP}}},\ }\href@noop
	{} {\enquote {\bibinfo {title} {{FMC10x} user manual},}\ }\bibinfo
	{howpublished} {\url{http://www.4dsp.com/FMC107.php}} (\bibinfo {year}
	{2011}{\natexlab{a}})\BibitemShut {NoStop}%
	\bibitem [{\citenamefont {{4DSP}}(2011{\natexlab{b}})}]{DS_FMC116}%
	\BibitemOpen
	\bibfield  {author} {\bibinfo {author} {\bibnamefont {{4DSP}}},\ }\href@noop
	{} {\enquote {\bibinfo {title} {{FMC116/FMC112} user manual},}\ }\bibinfo
	{howpublished} {\url{http://www.4dsp.com/FMC116.php}} (\bibinfo {year}
	{2011}{\natexlab{b}})\BibitemShut {NoStop}%
	\bibitem [{\citenamefont {Heinzel}\ \emph {et~al.}(2011)\citenamefont
		{Heinzel}, \citenamefont {Esteban}, \citenamefont {Barke}, \citenamefont
		{Otto}, \citenamefont {Wang}, \citenamefont {Garcia~Marin},\ and\
		\citenamefont {Danzmann}}]{Heinzel2011}%
	\BibitemOpen
	\bibfield  {author} {\bibinfo {author} {\bibfnamefont {G.}~\bibnamefont
			{Heinzel}}, \bibinfo {author} {\bibfnamefont {J.~J.}\ \bibnamefont
			{Esteban}}, \bibinfo {author} {\bibfnamefont {S.}~\bibnamefont {Barke}},
		\bibinfo {author} {\bibfnamefont {M.}~\bibnamefont {Otto}}, \bibinfo {author}
		{\bibfnamefont {Y.}~\bibnamefont {Wang}}, \bibinfo {author} {\bibfnamefont
			{A.~F.}\ \bibnamefont {Garcia~Marin}}, \ and\ \bibinfo {author}
		{\bibfnamefont {K.}~\bibnamefont {Danzmann}},\ }\href@noop {} {\bibfield
		{journal} {\bibinfo  {journal} {Classical and Quantum Gravity}\ }\textbf
		{\bibinfo {volume} {28}},\ \bibinfo {pages} {094008} (\bibinfo {year}
		{2011})}\BibitemShut {NoStop}%
	\bibitem [{\citenamefont {Delgado}(2012)}]{Esteban2012}%
	\BibitemOpen
	\bibfield  {author} {\bibinfo {author} {\bibfnamefont {J.~J.~E.}\
			\bibnamefont {Delgado}},\ }\emph {\bibinfo {title} {Laser ranging and data
			communication for the laser interferometer space antenna}},\ \href@noop {}
	{Ph.D. thesis},\ \bibinfo  {school} {Leibniz Universität Hannover} (\bibinfo
	{year} {2012})\BibitemShut {NoStop}%
	\bibitem [{\citenamefont {Sutton}\ \emph {et~al.}(2013)\citenamefont {Sutton},
		\citenamefont {McKenzie}, \citenamefont {Ware}, \citenamefont {de~Vine},
		\citenamefont {Spero}, \citenamefont {Klipstein},\ and\ \citenamefont
		{Shaddock}}]{Sutton2013}%
	\BibitemOpen
	\bibfield  {author} {\bibinfo {author} {\bibfnamefont {A.~J.}\ \bibnamefont
			{Sutton}}, \bibinfo {author} {\bibfnamefont {K.}~\bibnamefont {McKenzie}},
		\bibinfo {author} {\bibfnamefont {B.}~\bibnamefont {Ware}}, \bibinfo {author}
		{\bibfnamefont {G.}~\bibnamefont {de~Vine}}, \bibinfo {author} {\bibfnamefont
			{R.~E.}\ \bibnamefont {Spero}}, \bibinfo {author} {\bibfnamefont
			{W.}~\bibnamefont {Klipstein}}, \ and\ \bibinfo {author} {\bibfnamefont
			{D.~A.}\ \bibnamefont {Shaddock}},\ }\href@noop {} {\bibfield  {journal}
		{\bibinfo  {journal} {Class. Quantum Grav.}\ }\textbf {\bibinfo {volume}
			{30}},\ \bibinfo {pages} {075008} (\bibinfo {year} {2013})}\BibitemShut
	{NoStop}%
\end{thebibliography}

\end{document}